\documentclass[12pt,preprint]{aastex}

\begin{document}


\title{Wall emission in circumbinary disks: \\
    the case of CoKu Tau/4}

\author{Erick Nagel\altaffilmark{1,2}, Paola D'Alessio\altaffilmark{1}, Nuria Calvet\altaffilmark{3}, Catherine Espaillat\altaffilmark{4,5}, Ben Sargent\altaffilmark{6}, Jes\'us Hern\'andez\altaffilmark{7}, and William J. Forrest\altaffilmark{8}}
\affil{Centro de Radioastronom\'\i a y Astrof\'\i sica, UNAM, Morelia,
       Michoac\'an, M\'exico 58089}
\affil{Departamento de Astronom\'\i a, Universidad de Guanajuato, Guanajuato,
       Gto, M\'exico 36240}
\affil{Department of Astronomy, University of Michigan, Ann Arbor,MI 48109}
\affil{Harvard-Smithsonian Center for Astrophysics, 60 Garden
Street, MS-78, Cambridge, MA, 02138} 
\affil{NSF Astronomy \& Astrophysics Postdoctoral Fellow} 
\affil{Space Telescope Science Institute, Baltimore, MD 21218}
\affil{Centro de Investigaciones de Astronom\'\i a, M\'erida 5101-A, Venezuela}
\affil{Department of Physics and Astronomy, University of Rochester, Rochester,
NY 14627}
\email{erick@astro.ugto.mx}


\begin{abstract}
A few years ago, the mid-IR spectrum of a Weak Line T Tauri Star, CoKu Tau/4,
 was explained as emission 
from the inner wall of a {\it circumstellar} disk, with the inner disk truncated at 
 $\sim$ 10 AU. Based on the SED shape and the assumption that it was produced
by a single star and its disk, CoKu Tau/4 was classified as a prototypical
transitional disk, with a clean inner hole possibly carved out by a planet, some other
orbiting body, or by photodissociation.
 However, recently it has been discovered that CoKu Tau/4 is a close binary 
system. This implies that the observed mid-IR SED is probably produced 
by the {\it circumbinary} disk. 
The aim of the present paper is to model the SED
of CoKu Tau/4 as arising from the inner wall of a circumbinary disk, with parameters constrained 
by what is known about the central stars and by a dynamical model for the 
interaction between these stars and their surrounding disk. 
We lack a physical prescription for the shape of the wall, thus, here we use a
simplified and unrealistic assumption: the wall is vertical.
In order to fit the Spitzer IRS SED, 
the binary orbit should be almost circular, implying a small mid-IR 
variability (10 \%) related to the variable distances
of the stars to the inner wall of the circumbinary disk. 
In the context of the present model, higher eccentricities would imply 
that the stars are farther from the wall, 
the latter being too cold to explain the observed SED.
Our models suggest that the inner wall of CoKu Tau/4 is located at 
$1.7 a$, where
$a$ is the semi-major axis of the binary system ($a\sim 8AU$). 
A small amount of optically thin dust in the hole ($\lesssim 0.01$ lunar masses) helps to improve the fit to the 10 $\mu$m silicate band. 
Also, we find that water ice should be absent or have a 
very small abundance (a dust to gas mass ratio $\lesssim 5.6 \times 10^{-5}$). 
In general, for a binary system with eccentricity $e > 0$, the model predicts
mid-IR variability with periods similar to orbital timescales, assuming that 
thermal equilibrium is reached instantaneously. 
\end{abstract}


\section{Introduction}

Until recently, the Weak Line T Tauri star, 
CoKu Tau/4 \citep{Cohen}, was known as a remarkable
example of a transitional disk \citep{Forrest}. Its mid-IR SED 
(IRS Spitzer) was modeled by \citet{Dalessio2}, hereafter D05, taking an 
isolated star, and they
concluded that the disk surrounding CoKu Tau/4 should have a clean 
inner hole, with a radius about 10 AU. In this model, the emergent 
flux in the mid-IR spectral range was produced by the inner wall of the 
truncated disk.  

Previously, it had been thought that the inner hole of CoKu Tau/4  was due to
either a recently formed planet \citep{Quillen} or by the UV switch model 
\citep{Clarke,Alexander} due to its undetectable accretion signatures. 
However, observations described by 
\citet{Ireland} show conclusively that CoKu Tau/4 is a binary
system, implying that its inner hole is probably produced by dynamical
 interactions between the binary system and the surrounding gas.

For a binary system formed by stars of similar masses 
 and using the predictions of \citet{Artymowicz1} and 
\citet{Beust}, \citet{Ireland} estimated that the  
size of the hole in the circumbinary disk (CB hereafter) should be between 13 
and 16 AU, for different eccentricities of the binary orbits. 
The estimated hole is larger than the one found in the D05 transitional disk 
model, 
and being further away, one would naively expect it to be cooler.
However, the stars are not at the geometrical center of the disk: each
point in the visible part of the wall is at a different distance from the 
stars. Therefore, the net effect of both stars in 
the heating of the wall should be calculated for a particular configuration 
before comparison with the observed SED.


This paper has several goals. The first is to find a simplified structure
of the inner edge of the CB disk around the
CoKU Tau/4 binary system that consistently reproduces the observed
spectral energy distribution (SED). The second is to determine the properties
of the dust particles of that edge. Spitzer observations probe mostly the
upper layers of the inner 10-20 AU regions of typical disks 
\citep{Dalessio3}, while
at the present resolution of submillimeter and millimeter interferometers,
the main contribution comes from the midplane at larger distances. In contrast, 
circumbinary and transitional disks are truncated at a certain radius, making 
them a natural laboratory
to study the dust near the midplane in the inner disk region, inaccessible 
by other means. In \S~\ref{sec-description} we describe the physics of the 
models, and in \S~\ref{sec-fiducial} we present the fiducial model adopted as
a reference for later comparisons; in \S~\ref{sec-var-param} we study
the effects of changing parameters in the SED of the fiducial model; in 
\S~\ref{sec-results} we apply our methods to the case of CoKu Tau/4, and in 
\S~\ref{sec-conclusions} we discuss our results and give conclusions.

\section{Description of the model}
\label{sec-description}


Our basic model consists of a CB disk, from which only 
its inner wall contributes to the SED in the mid-IR spectral range 
($5-40\mu$m), surrounding two stars that contribute mainly to the near-IR flux.
The main difference between the present model and the one described in D05 is
that here the inner wall is irradiated by two stars and each 
of them has a 
different distance to any surface element of the wall since
neither of the stars is located at the geometrical center of the CB 
disk.  Moreover, since the stars are moving in their orbits, their distance 
to each portion of the inner wall changes with time. Assuming that the wall
reaches thermal equilibrium in a timescale shorter than the orbital timescale
of the binary system (see the Appendix for the evaluation of this assumption), 
the temperature of each portion of the wall 
and the resulting SED are periodically variable with
its orbital timescale. 

The fact that we are describing a CB disk implies that 
a whole set of new parameters should be included into the model, in 
comparison to the model proposed by D05. 
At first sight, it might look like this increasing  
degree of freedom  makes the problem degenerate. 
However, new restrictions allow one to break part of this degeneracy, as we will
see in the following sections.  

\begin{figure}
\epsscale{.80}
\plotone{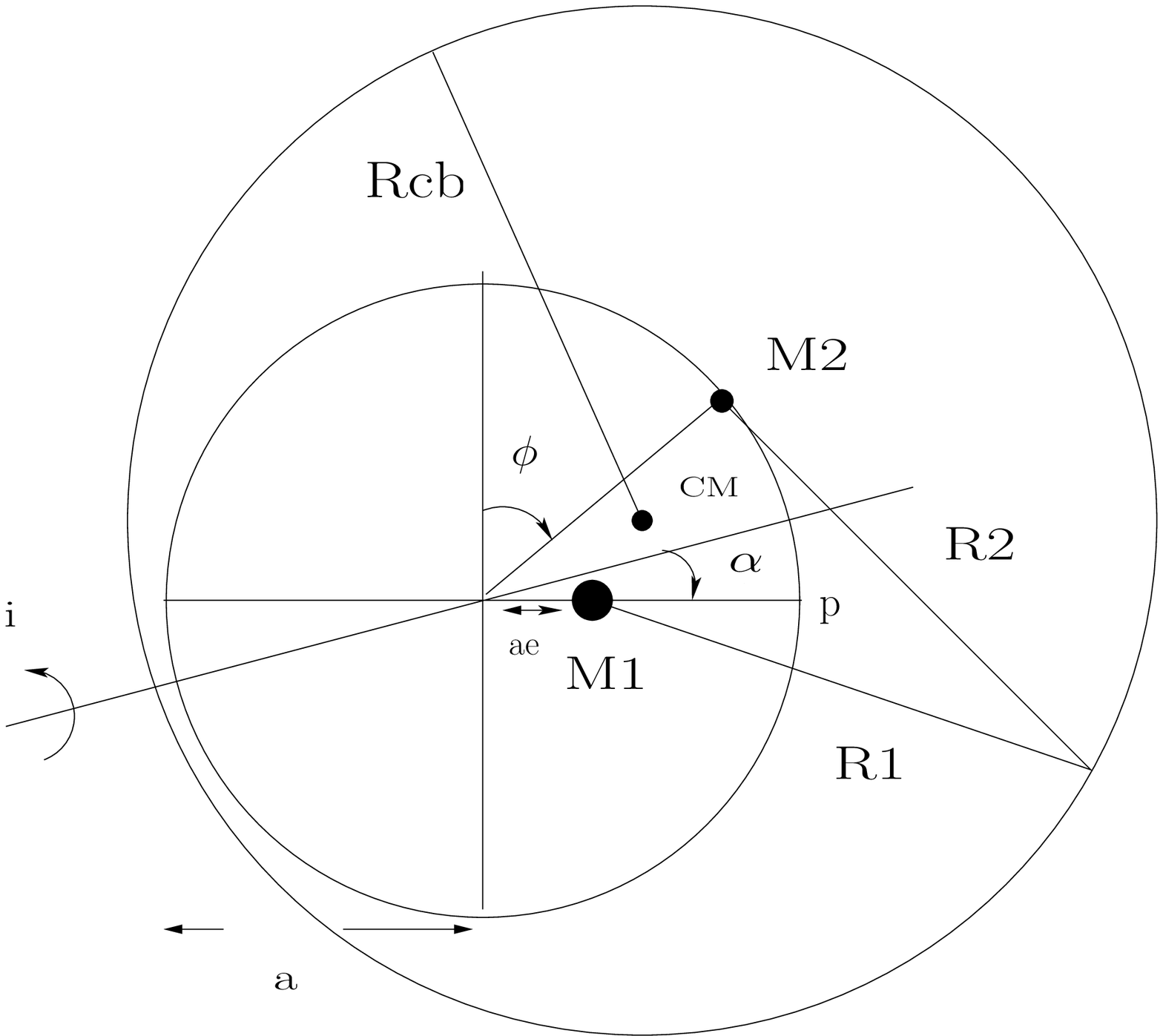}
\caption{Schematic diagram of the parameters that characterizes the system 
         configuration in space. The parameters shown are $M1$,$M2$,$R1$,
         $R2$,$R_{cb}$,$\alpha$,$\phi$,$i$,$a$ (see text for definitions),$ae$ 
         is the distance between the center of
         the ellipse and the star in the focus, $p$ corresponds to the 
         pericenter, the stars configuration of closest approach. The locations
         of the stars are given with respect to the center of the ellipse, and the center
         of the CB disk is given with respect to the center of mass (CM) of the 
         binary system. The line given by the angle $\alpha$ is the one around
         which the system is inclined at an angle $i$.
         The figure is not to scale. See text for more details.}
\label{fig-configuracion}
\end{figure}

\begin{figure}
\scalebox{1.0}[0.8]{
\plotone{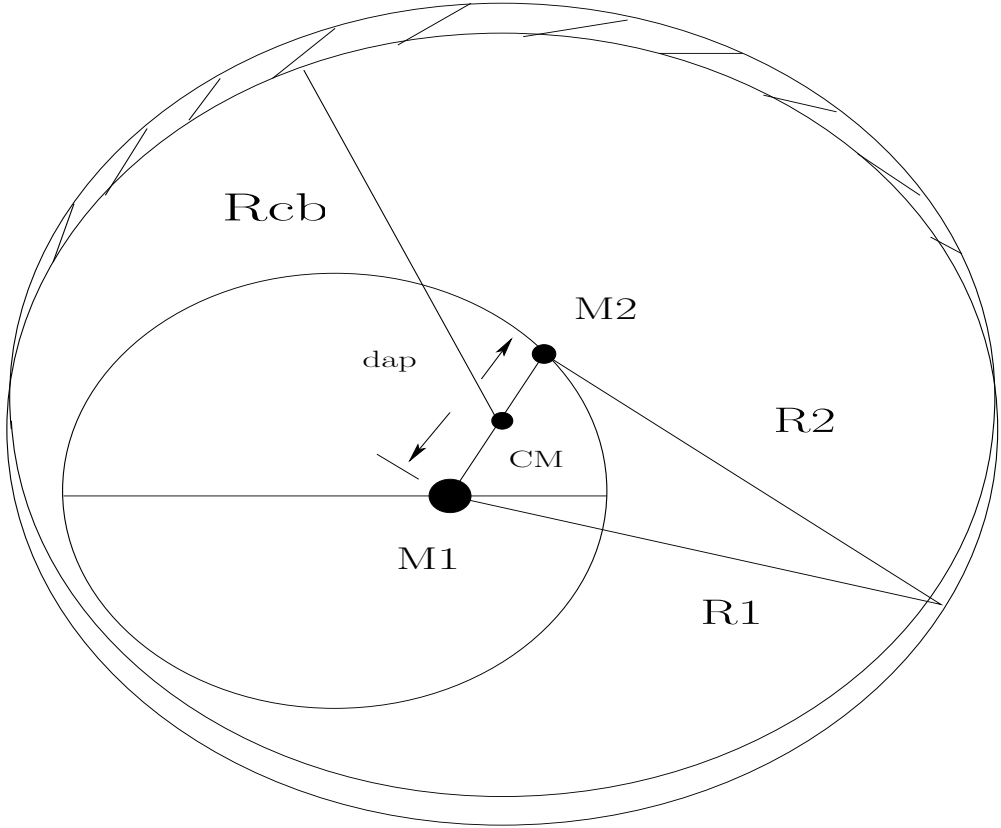}}
\caption{Schematic diagram of an inclined system with arbitrary $i$. 
  In this example, the system is inclined around the 
  major axis line, $\alpha=0.$ Here, $d_{ap}$ corresponds to the distance 
  between the stars in the plane of the sky. 
  The dashed area shows the fraction of the wall directed towards the observer.
  The rest of the parameters are described in the text. The only difference
  between Figure~\ref{fig-configuracion} and this figure is that this one 
  corresponds to an inclined system. The figure is not to scale.}
\label{fig-configuracion-i}
\end{figure}

Figure~\ref{fig-configuracion} shows the geometrical parameters that describe 
the spatial configuration of the system, namely,
the semi-major axis $a$, eccentricity $e$ and the angle $\phi$, that 
locates one of the stars in an elliptical orbit with the other one at the 
focus. Figure~\ref{fig-configuracion} shows the two angles that are required 
to characterize this configuration in 
3D space: $i$, the inclination between the disk axis and the line of sight
(or, between the disk plane and the plane of the sky) and $\alpha$, which gives
the axis in the disk plane around which the disk is inclined. This line passes 
through the center 
of the elliptical orbit and the angle $\alpha$ is measured between this line 
and the major axis line. Note that first the system is described in the 
orbital plane with $a$, $e$ and $\phi$ (see Figure~\ref{fig-configuracion}).
Then, the major axis line is 
identified, from which an angle $\alpha$ gives the line around which the system
is inclined an angle $i$. An inclined system is shown in 
Figure~\ref{fig-configuracion-i}. For these parameters, the origin of the 
reference system is in the center of the ellipse.
Notice that changing $\alpha$  not only changes the orientation 
of the visible part of the wall, but also modifies the distances 
from the stars to
that part of the wall (\S~\ref{sec-wall-emission}). On the other 
hand, a variation in $i$ changes the visible area of the
wall (\S~\ref{sec-wall-emission}).   
In the following, the angles 
$\phi$ and $\alpha$ are given in units of $\pi$ radians.

The inner wall or inner boundary of the CB disk is 
characterized by its radius $R_{cb}$, and its vertical height, $h$. 
Notice that in this system of reference, 
while the second star describes the elliptical orbit, the
center of mass of the binary system and also the center of the 
hole in the CB disk move (both points are the same for $e=0$ but not for other 
values of eccentricity, e.g., \citet{Pichardo1} and \citet{Pichardo2}). Thus, 
during a 
binary system orbit, the distance between the two stars and the wall changes.
For the following, the origin of our system of reference is the center of the 
CB disk. 

Also, it is important to specify the dust 
composition of the material in the hot atmosphere of the wall. 
We assume that, in general, the dust is a mixture of silicates, troilite,
graphite, amorphous carbon and water ice. The silicate (olivine or pyroxene) 
and graphite abundances are given by \citet{Draine2}. The maximum abundance
of water ice and the abundance of troilite is taken from \citet{Pollack}.
We change the dust 
composition and abundances in \S~\ref{sec-var-param} and adjust them by 
comparing the synthetic and the observed SEDs in \S~\ref{sec-results}.

A constraint inherent to a two-star model is given by
the dynamical interaction of the binary system with a disk. 
This interaction clears an inner hole, thus, in this 
way the radius $R_{cb}$ is determined by the stars' masses and the
eccentricity and semi-major axis of their orbits. 
In contrast, in the case of one star (D05) $R_{wall}$ is a free parameter. In
order to reproduce the observed SED, $R_{wall}$ was adjusted to obtain the 
appropriate temperature structure.

The size of the hole ($R_{cb}$) can be calculated numerically (\citet{Rudak}, 
\citet{Pichardo1},\citet{Pichardo2}, \citet{Artymowicz1}) or 
analytically for a circular binary \citep{Nagel}. However,  
\citet{Artymowicz1} take a hydrodynamical 
approach with the inclusion of viscosity, in addition to the gravitational 
interaction
of the stars and the disk material. The presence of viscosity allows the
material to move to the region where there are no stable orbits around both 
stars. If a mass particle passes a certain point, then the only possibility 
for it is to move to one of the stars. In this way, streams of material 
originating in
the CB disk connect to the circumstellar disks (these are disks inside the hole
surrounding each star), acting as a feeding mechanism
\citep{Artymowicz2,Gunther1}. Observations by 
\citet{Jensen} suggest the presence of streams from the CB disk towards the 
binary 
system UZ Tau E. Also, for the binary St 34, \citet{Hartmann} argue 
that the 
presence of dust in the dynamically cleared hole is consistent with having some
accretion flow onto the central binary. \citet{Gunther2}, 
described the emission of DQ Tau and AK Sco using a hydrodynamical simulation,
where all the disks and the streams connecting them are taken into account.
Also, binaries of very low mass are expected to be fed by a circumbinary disk
\citep{Robberto}.
From all these studies, we can argue that the truncation radius $R_{cb}$ 
strongly
depends on the viscosity in a non trivial way. For an $e>0$
system, \citet{Artymowicz1} show clear differences between $R_{cb}$ for
changes in the Reynolds number, which depends on the viscosity. However, in
order to take a more realistic model, we decided to use the results of 
\citet{Artymowicz1} taken from Figure 3, arguing that $R_{cb}$ does not
depend strongly on the stars mass ratio, and a Reynolds number of $10^{4}$. 


An important ingredient for the calculation of the radiation flux from the 
stars impinging on the circumbinary wall is the distance between each star and
a specific point on the wall, $R_{1,2}$ (see Figures~\ref{fig-configuracion} 
and \ref{fig-configuracion-i}). In order to calculate these distances, 
the first step is to give the locations of the stars
in the configuration where one star is at the focus and the other is 
on a point on its corresponding ellipse.
The second step is to place a circular CB disk with its origin at the point
corresponding to the eccentricity of the binary system 
\citep[see][]{Pichardo1,Pichardo2}, e.g., the center of mass of the binary 
system if $e=0$,
with a radius given by \citet{Artymowicz1}.  Finally, 
the third step is to calculate the distances between the coordinates of the 
stars and each surface element of the  wall given in this reference system
(see Figures~\ref{fig-configuracion} and \ref{fig-configuracion-i}).  

\subsection{Formulation of the wall emission problem}
\label{sec-wall-emission}

The inner wall of the CB disk is assumed to
be vertical, i.e., the vector normal to its surface is parallel to the
disk midplane. In order to relax the assumption of verticallity one requires a way to describe the shape of the wall, which is not
possible to the best of our knowledge, because the observations and simulations do not have 
the necessary resolution. More complicated shapes have been assumed for the
dust to gas transition at the dust destruction radius in the inner disk by
taking into account the dependence of the sublimation temperature or density
\citep{Isella} or the maximum grain size \citep{Tannirkulam}.
However, the edge of the CB disk is a physical wall created by tidal 
interactions with the binary system, and the vertical density profile may be 
very different from that in standard disks. Since no detailed models exist to
guide us, we have adopted the simplest geometry: a vertical wall.  The 
wall is also assumed
to be circular for simplicity, however, non-axisymmetric shapes are expected
based on hydrodynamical simulations of \citet{Gunther1}
and \citet{Artymowicz2}. In Section~\ref{sec-var-param-wall} we show 
that the SED for an eccentric CB disk does not change appreciably with respect
to the circular case, thus the CoKu Tau/4 modeling is restricted to the latter. 

 Only a fraction of the wall's surface can be detected by the
observer, depending on the wall radius, height and the inclination
angle $i$. There are other 
portions of the wall that are occulted from the observer by absorption due to the
disk itself. The ``visible'' surface, projected in the plane of the sky (see 
Figure~\ref{fig-configuracion-i}), is divided in small surface
elements (pixels).  Each one is located at a given height relative to the disk
midplane, $Y$ and at a given distance, $X$, from the disk center. In order to
describe each pixel position, we use cartesian coordinates (see D05).

Each point in the wall has a different temperature distribution, from
the surface toward larger radii (deeper inside the disk). In order to
estimate this temperature distribution we use a similar treatment
given by \citet{Calvet2} and \citet{Calvet1}.  However, in the present
case each pixel is being irradiated by two stars, neither of which is
at the center of the system. We use the fact that the luminosity from the wall 
is much less than the luminosity from the stars in order to neglect its 
contribution to the heating of the wall. Thus, at each pixel we consider the
superposition of two radiation fields, characterized by two different
fluxes and mean intensities, and, in principle, by two distinct
incidence directions. The incident flux from each star
is $F_{0,j}= ({L_{*,j}/4 \pi R_j^2})\mu$ with each star at a different
distance from the wall, given by $R_j$, $\mu$ is the cosine of the incidence 
angle with respect to the normal to the wall surface. Again,
for the sake of simplicity, we assume that each pixel is a
plane parallel atmosphere and the radiation from each star arrives in
a single beam, with a normal incidence angle ($\mu=1$).  This assumption is
valid only if the minimum distance between the stars and the wall is much
larger than the stellar radii. In the case of CoKu Tau/4, the minimum
distance between one star and the wall is 1.2 $a \sim 9 AU$, so the
later assumption is justified for this particular object.

Following \citet{Calvet2} and D05, the radiation field arriving at
each pixel is then separated into three components: ``stellar 1'',
``stellar 2'' and ``disk'', each one with a particular wavelength
range where most of its emission peaks. Defining the optical depth at
the ``disk'' wavelength range in the radial direction, measured from the wall
surface towards larger radii, the net outward Eddington flux at the
disk frequency range, $H_d$, is given by

\begin{equation}
 H_d(\tau_d)= \sum_{j=1}^2 \frac{F_{0,j}}{4 \pi} \alpha_1
[(1+C_{1,j}) e^{-q_j \tau_d} + C_{2,j} e^{-\beta_j q_j \tau_d} ] ,
\label{eq_hd} 
\end{equation}

\noindent 
where $\alpha_j$ is the mean true absorption coefficient, $\kappa_{s,j}$,
 divided by
the mean total disk opacity, $\chi_d$ 
(calculated including true absorptions and scattering), 
evaluated using the effective temperature of each star (given by $j=1$
or $2$), and  $\beta_j=(3\alpha_j)^{1/2}$. Here, $q_j(= \chi_{s,j}/\chi_d$),
 is the ratio between
the mean total opacity at the stellar wavelength range and the mean
total opacity at the disk range (which is the opacity used to
calculate $\tau_d$).  The constants in equation (\ref{eq_hd}) are
given by equations 2 and 3 in D05, but in here they are evaluated
using the mean opacities corresponding to each star. 

All the mean opacities are
calculated using the Planck function evaluated at the local or at
the stellar effective temperature as the weighting function. The reason is
that this kind of average is easy to calculate and the mean contributions of
scattering and true absorption can be added together to give the total opacity,
in contrast to a Rosseland kind of average \citep{Dalessio1}. Using this 
kind of mean opacity we
obtain the correct value of temperature at $\tau_{d}\rightarrow 0$, because 
this is adequate for the optically thin layers of the wall atmosphere 
\citep{Mihalas}.

The Eddington flux leads to the mean intensity at the disk
frequency range, using the first momentum of the transfer equation and
the usual boundary condition $J_d(0)=2H_d(0)$. 

From the zeroth
moment of the transfer equation, assuming isotropic scattering, 
the temperature can be written as,

\begin{equation}
T(\tau_d)^4= \sum_{j=1}^2 \alpha_j \frac{F_{0,j}}{4 \sigma_R}(C_{1,j}^\prime 
+ C_{2,j}^\prime e^{-q_j \tau_d} + C_{3,j}^\prime e^{-\beta_j q_j \tau_d}), 
\label{eq_t}
\end{equation}

\noindent
with constants given in D05. Another, more compact, 
way to write down equation (\ref{eq_t}) 
is

\begin{equation}
T(\tau_d)^4= T_1(\tau_d)^4 + T_2(\tau_d)^4,
\end{equation}

\noindent
where $T_1$ and $T_2$ are the temperatures calculated as if only the
star $1$ or $2$ are present, respectively, as a superposition
solution.  However, in spite of its apparent simplicity, this is
really an implicit equation for the temperature since the mean
opacities involved in the calculation of $T_1$ and $T_2$
depend on the temperature $T$.

As in \citet{Calvet2}, \citet{Calvet1} and D05, we have assumed that the 
opacities
are constant in the wall atmosphere, i.e., $\alpha_j$ and $q_j$ are
independent of $\tau_d$. If the opacity is dominated by dust, and the
temperature distribution in the wall atmosphere is such that there is
no sublimation of any dust ingredient at a particular optical depth,
then it is valid to assume that $\alpha_j$ is constant. Also, if
the contrast in temperature between the uppermost and the deepest
layers is not very high,  $\chi_d$ and $q_j$ can be assumed to be
constant too. 
Thus, in order to quantify the mean opacities, we use the temperature
evaluated at $\tau_d=0$, i.e., the surface temperature $T_{wall}$. This is
found solving the implicit equation (\ref{eq_t}), for a given wall
radius $R_{cb}$, a given configuration of the binary system, and the 
coordinates of each pixel in the wall, a pair  $(X,Y)$.

The monochromatic emergent intensity from each pixel of the wall 
in the direction of
the observer is calculated by integrating the transfer equation. 
We consider two components: the thermal emission 
and the scattering of stellar radiation.  Finally, the SED is
calculated by multiplying the emergent intensity by the solid angle
subtended by each (``visible'') pixel, as seen by the observer.

\subsection{Formulation of the hole emission problem}
\label{sec-gap-emission}

As described in \S~\ref{sec-description}, the size of the hole ($R_{cb}$)
is given by the dynamical interaction between the disk and the stars. 
However, the
full hydrodynamical treatment should include ingredients like 
the gas  viscosity, that 
allows material from the CB disk to lose angular momentum and fall into 
the hole. This 
results in a small quantity of material inside the CB disk hole, 
that can be treated as being optically thin.
Avoiding the precise description of the dynamical
evolution of such particles, in this work we simplify the problem considering
them distributed in a region located between a radius $R=R_{min}$ and 
$R=R_{cb}$, uniformly distributed in the azimuthal angle in a cylindrical 
coordinate system, centered at the center of mass of the binary system.
Indeed, simulations of the material inside the hole 
\citep{Artymowicz2,Gunther1} indicate that the material is non axisymmetric 
distributed and spans all radii less than $R_{cb}$. However, for a circular
binary system with undetectable accretion towards the stars, we can argue that
most of the material is restricted to lie at radius larger than the outer
Lagrangian point for the three body circular restricted problem 
\citep{Szebehely}.
The surface density is given by a power law ($\Sigma\propto R^p, p=1$) such 
that $\Sigma$ increases closer to the wall, as is expected 
\citep{Artymowicz2}.  

The emergent flux from the dust in the hole depends on the dust temperature.
Since we assume the hole is optically thin, then each dust grain is heated 
by absorption of geometrical diluted  radiation from each star and 
cools down by emission of its own thermal radiation. This is expressed
by the radiative equilibrium equation, written as

\begin{equation}
\int_0^\infty \kappa_\nu J_{\nu}^* d\nu = \int_0^\infty \kappa_\nu B_{\nu}(T(R,\theta)) d\nu, 
\end{equation}

\noindent
where the mean intensity of the radiation field is approximated by

\begin{equation}
J_{\nu}^* \approx W_1(R_1, R_{*1}) B_\nu(T_{*1})+ W_2(R_2, R_{*2}) B_\nu(T_{*2}),
\end{equation}

\noindent
with the dilution factors, 

\begin{equation}
W_j = \frac{1}{2} \biggl [ 1 - \sqrt{ 1 - \biggl ( \frac{R_{*j}}{R_j} \biggr )^2} \biggr ], {\rm for \ j=1,2}.
\end{equation}

When the integrals are performed, the dust grain temperature is the solution
of the implicit equation

\begin{equation}
T(R,\theta)^4 =  \sum_{j=1}^2 W_j T_{*j}^4 [\kappa_{s,j}/\kappa_d(T(R,\theta))],
\end{equation}

\noindent
where $\kappa_d(T(R,\theta))$ is the Planck mean opacity given by

\begin{equation}
\kappa_d(T(R,\theta))= {\int_0^{\infty} \kappa_{\nu} (T) B_{\nu}(T) d\nu\over \sigma T^4/\pi},
\end{equation}

\noindent
and $\kappa_{s,j}(T(R,\theta),T_{*j})$ is a Planck-like mean opacity, which can
be written as

\begin{equation}
\kappa_{s,j}(T(R,\theta),T_{*j})={\int_0^{\infty} \kappa_{\nu} (T) B_{\nu}(T_{*j}) d\nu\over \sigma T_{*j}^4/\pi}.
\end{equation}

Here, we use the dust temperature to evaluate the monochromatic opacity and the
stellar effective temperature (for each star, $j=1$ or $2$) to evaluate the 
Planck function in the average. The Planck-like mean opacity is the 
straightforward result from the solution of the radiative equilibrium equation.
 
The monochromatic opacities we use 
are calculated assuming a size distribution for the grains. 
In the present work, which is our first approximation to the problem, 
we do not calculate a particular  
temperature for each grain size, but a mean temperature for the whole size
distribution. In this approximation, the emission is produced from small to 
large grains at the same temperature, however, for the full case, the emission 
from small and large grains will differ due to their different temperature. 
Thus, we assume that the differences in the emitted spectrum between the 
approximation and the real case are small. 
In \S~\ref{sec-var-param} we explore different ingredients for the dust in the 
hole.



\section{Fiducial model}
\label{sec-fiducial}

As described, this problem is too complex. Moreover, a full analysis of all the 
parameters involved is beyond our actual computational resources. Thus, we 
restrict the problem to variations around a fiducial model (E1 in 
table~\ref{table-models}), where some parameters
are fixed. 
Observations of the stellar parameters and orbital parameters help
restrict the set of free parameters. Comparison between the
synthetic and observed SED then provides a range for the rest of
the parameters.


We adopt parameters relevant to CoKu Tau/4 for the fiducial model.
First of all, the model requires the parameters that characterize the stars. 
What is assumed for the stars is mostly based on  the observations 
reported by \citet{Ireland} for CoKu Tau/4. The 
masses of the stars ($M_{\star 1}$ and $M_{\star 2}$) have a ratio 
$M_{\star 2}/M_{\star 1}= 0.85 \pm 0.05$, and near-IR colors consistent 
with typical T Tauri spectral types.  These stellar masses, 
combined with $a$, give the distance
of each of the stars to the center of mass of the binary system. 
Also, assuming the stars are coeval, by specifying each of their masses, their 
ages and a stellar evolutionary model, 
one obtains the stellar radii  ($R_{\star 1,2}$), 
their effective temperatures ($T_{\star 1,2}$) and their luminosities 
($L_{\star 1,2}$).  

In this paper we adopt the stellar
evolutionary tracks from \citet{Siess}, and the spectra are
taken from synthetic stellar spectrum from libraries from A.~Bruzual, which are
described in \citet{Bruzual}. For practical purposes, we take
the sum of the spectra of two appropriate stars for comparison with the 
observed spectra at large frequencies. This process results in
two spectral types for both stars. \citet{Ireland} propose the stars 
have $M1.5\pm 0.5$ spectral types, consistent with an age of $4 \ Myrs$. 
Here, we use the evolutionary tracks from \citet{Siess} given in
$0.1M_\odot$ steps. Taking this restriction, we adopt $M0$ and 
$M1$ spectral type stars, which are within the uncertainties (A. Kraus,
personal communication). Table~\ref{table-stars} shows the parameters for 
the stars. 

\begin{figure}
\epsscale{.80}
\plotone{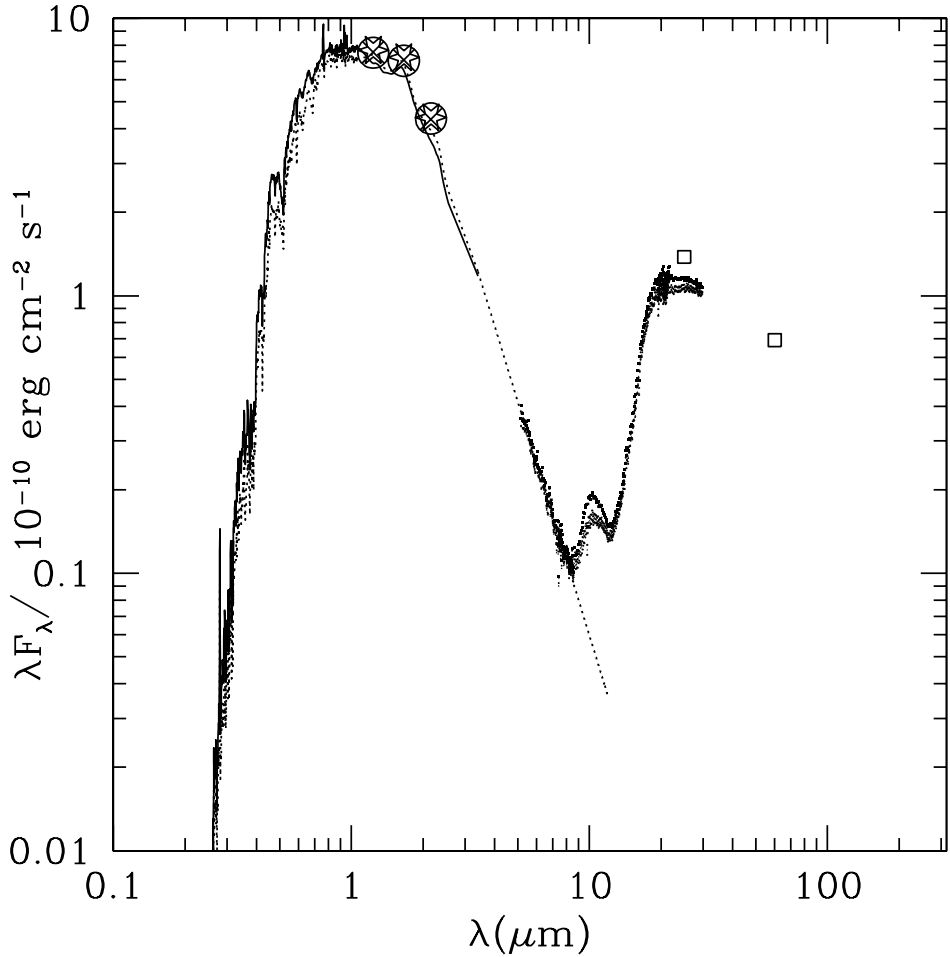}
\caption{Combined spectrum of the central stars for a $4 \ Myr$ old binary
system (solid line). The observed mid-IR SED of CoKu Tau/4 is corrected by 
four different extinction laws: \protect\citet{Moneti} (upper curve), 
\protect\citet{McClure} 
(next curve), \protect\citet{Draine1} (next curve) and \protect\citet{Mathis2} 
(lower curve). 
The three last curves are very similar. 
IRAS data \protect\citep{Furlan} are represented by open squares,
and the 2MASS data are crosses, circles
and stars, corresponding to Mathis, Moneti and Draine extinction laws, 
respectively.
The synthetic spectrum adopted in D05 is also 
plotted (dotted line).}
\label{fig-stars-spectrum}
\end{figure}

Figure~\ref{fig-stars-spectrum} shows the stellar spectra for the 
binary stars with an age of $4 \ Myrs$, corrected using 
$A_{V}=3$ and four different extinction laws 
\citep{Moneti,Draine1,McClure,Mathis2}. The spectra corrected with 
\citet{Draine1},\citet{Mathis2}, and \citet{McClure} are very close. Note that 
the extinction law in the near-IR does not introduce noticeable changes in the
SED, but it affects the mid-IR spectral region 
(see Figure~\ref{fig-stars-spectrum}).

We note that 
from the spectrum there is no evidence of circumstellar disks, since the SED of
CoKu Tau/4 at $\lambda \leq 8\mu m$ seems to be completely 
described by photospheric emission of the stars \citep{Ireland}.
Moreover,  CoKu Tau/4 is classified a WTTS, with an $H\alpha$ equivalent width 
consistent with negligible accretion
from any possible circumstellar disks to the stars, 
with a rate constrained to be  $\dot{M} < 10^{-12}M_{\odot}/year$, which is an
observational threshold for accretion, according to \citet{Muzerolle}. 
Thus, in the fiducial model we do not consider the presence of circumstellar 
disks. 

We find that only a particular range
in wall temperatures [$T_{wall}(X,Y,\tau_d)$] produces the observed
SED consistent with the observed SED of CoKu Tau/4. $T_{wall}$ strongly depends
on the range of distances of the
stars to all the points in the wall. As noted above, $R_{cb}$ is no longer a 
free parameter and depends primarily on the eccentricity ($e$): 
if $e$ increases,  
$R_{cb}$ increases, too, and then both $T_{wall}$ and $F_{obs}$ decrease. For 
the fiducial model, we adopt the  \citet{Artymowicz1} value for $R_{cb}$ 
($=1.7a$) for a $e=0$ system for simplicity.

In order to fix $\alpha$ and $\phi$, we consider configurations with the 
highest emergent flux. 
For this to be done, we 
need to search between configurations along the whole orbit of the 
system. However, the angle $\phi$ is irrelevant because the orbit is circular,
but the angles $\alpha$ and $i$ have to be specified.
The former defines what fraction of 
the wall is directed towards the observer and the latter, 
the projection of the
wall in the plane of the sky. The stars are 
not illuminating the wall homogeneously and there are 
two intensity peaks in the positions (X,Y) closest to each of
the stars. 
The only configuration in which both intensity maxima are observed is the case
where the stars are located in the axis around which the system is inclined,
$\phi=0.5$ and $\alpha=0$.  
Thus, we choose this configuration in an effort to reduce the number of free
parameters and to get a reasonable model to start with. This also turns out to 
be the highest flux model. For the same reason, we take the minimum observed 
semi-major axis ($a=7.43AU$) as the fiducial value. As in D05, we choose 
$\cos i=0.5$ as a typical value.
In the fiducial model the height of the wall is assumed to be $h=0.28a$.

Finally, for the fiducial model the dust composition consists of graphite, 
silicates, troilite and no water ice, with dust to gas mass ratios
$\zeta_{grap}=0.0025$ and $\zeta_{sil}=0.0034$ \citep{Draine2}. The 
troilite abundance is $\zeta_{troi}=7.68\times 10^{-4}$ \citep{Pollack}. 
Graphite optical properties are taken 
from \citet{Weingartner} for graphite, and the silicates are assumed 
to be pyroxenes ($Mg_{0.8}\,Fe_{0.2}\,SiO_{3}$), with optical properties from 
\citet{Dorschner}. 
Also, we consider an interstellar medium dust size distribution 
\citep{Mathis1} with $n(a)\propto a^{-3.5}$, with minimum and maximum sizes 
$a_{min}=0.005\mu$m and $a_{max}=0.25\mu$m, respectively.

\section{Effect of parameters on SED}
\label{sec-var-param}


To show the dependence of the SED on the input parameters
we change one parameter at a time and compare the resultant SED to that of
the fiducial model. We first consider the change of parameters in the
wall, and then the change of parameters in the optically thin region
inside the hole. We assume that the apparent separation of the stars
on the plane of the sky is the same as the observed separation
$d_{ap}=53.6\pm 0.5mas = 7.5 \pm 0.07 AU$ \citep{Ireland}.
This assumption implies a fixed semi-major axis, $a$. Note that
the system configuration changes with time, thus, the $d_{ap}$ will change, for
a fixed $a$.

\subsection{Effects of wall parameters on SED}
\label{sec-var-param-wall}

In the modeling, the inner boundary of the CB disk is circular, which is the 
simplest assumption. However, in principle its shape can be non-axisymmetric.
From the results of the hydrodynamical simulations of \citet{Artymowicz2}
and \citet{Gunther1}, we note that in the line that connects 
both stars, the disk material is approaching the outer lagrangian points. Thus,
in order to quantify the difference between axisymmetric and non-axisymmetric
models, we model the shape of the wall as an ellipse with its semiminor axis 
along this line. This is an easy way to get a non-axisymmetric wall, however,
only a hydrodynamical simulation applied to systems like CoKu Tau/4 will allow
us to accurately characterize the wall. Thus, the wall presented here is just
one of many possibilities. 

\begin{figure}
\epsscale{.80}
\plotone{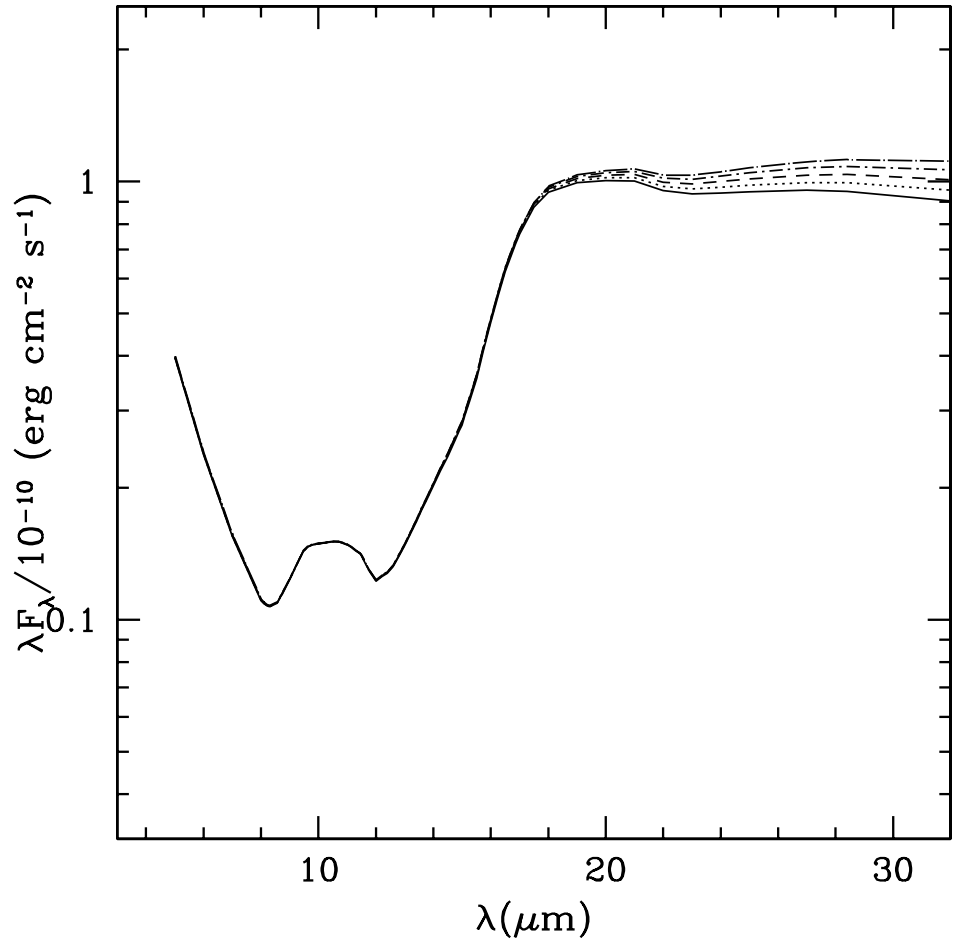}
\caption{SEDs of non-axisymmetrics models. The wall is an ellipse with a 
 semiminor
 axis (b) along the line connecting both stars. The binary system has an
 age of $4 \ Myrs$, with $\phi=0.5$,$\alpha=0$,$\cos i=0.5$, and $h=0.28a$.
 The models have $b=R_{cb}-0.4a$ (solid); $b=R_{cb}-0.3a$ (dotted);
	    $b=R_{cb}-0.2a$ (dashed); $b=R_{cb}-0.1a$ (dot-dashed), and
            $b=R_{cb}-0.0a$ (dot-long-dashed).}
\label{fig-noaxisim}
\end{figure}

Figure~\ref{fig-noaxisim} shows models with the parameters of the fiducial 
configuration, but with a semiminor axis for the CB disk of
$b=R_{cb}-(0.4,0.3,0.2,0.1,0.0)a$. Note that the flux decreases for smaller $b$,
which seems unphysical, due to the fact that the temperature of the wall 
closest to the stars is larger. However, the projected area perpendicular to
the line of sight decreases in such a way that the second effect dominates the 
observed emission.

The main result is that the changes are 
restricted to wavelengths longer than $12\mu$m, thus, the silicates band is not
modified. The changes are larger around $30\mu$m, but this is only relevant
for smaller $b$, which represents a highly eccentric disk.
This example shows how an elliptical wall modifies the SED, however, only when 
observations and/or simulations solve the structure of the wall, we will be able to quantify
this effect in the modeling of CoKu Tau/4.  

\begin{figure}
\epsscale{.80}
\plotone{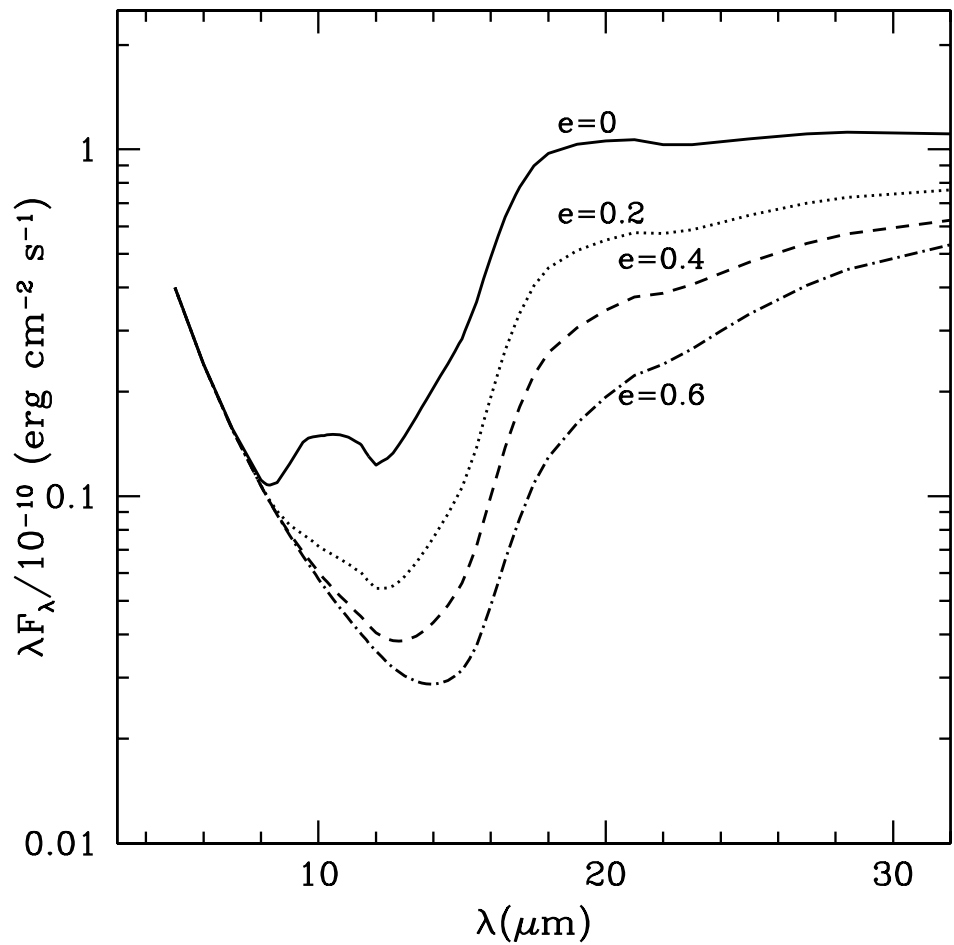}
\caption{SEDs of models with different binary eccentricities and an
age of $4 \ Myrs$, with $\phi=0.5$,$\alpha=0$,$\cos i=0.5$, and $h=0.28a$.  The
	    parameters are given in Table~\ref{table-models} for each model;
	    model E1 (solid); model E2 (pointed);
	    model E3 (dashed); and model E4 (dot-dashed).}
\label{fig-Draine-4Myr-exc}
\end{figure}

Now, we consider changes of the binary system eccentricity, $e$, for 
$\phi=0.5$,$\alpha=0$,
$\cos i=0.5$ and $h=0.28a$.
Figure~\ref{fig-Draine-4Myr-exc} shows the SED of the fiducial
model and of models with $e = 0., 0.2, 0.4,$ and $0.6$
calculated with $R_{cb}$ from \citet{Artymowicz1}. 
As $e$ increases, so does $R_{cb}$.  
This figure shows that
large values of $e$ imply SEDs from colder walls, farther away from the stars,
which show no 10 $\mu m$ silicate feature. The wall surface temperature 
varies from $136.5\,K$ for the 
smallest $R_{cb}$, and $80.7\,K$ for the largest $R_{cb}$. In
Table~\ref{table-models} we label these models as Series Ecc, where all the
parameters are shown.

\begin{figure}
\epsscale{.80}
\plotone{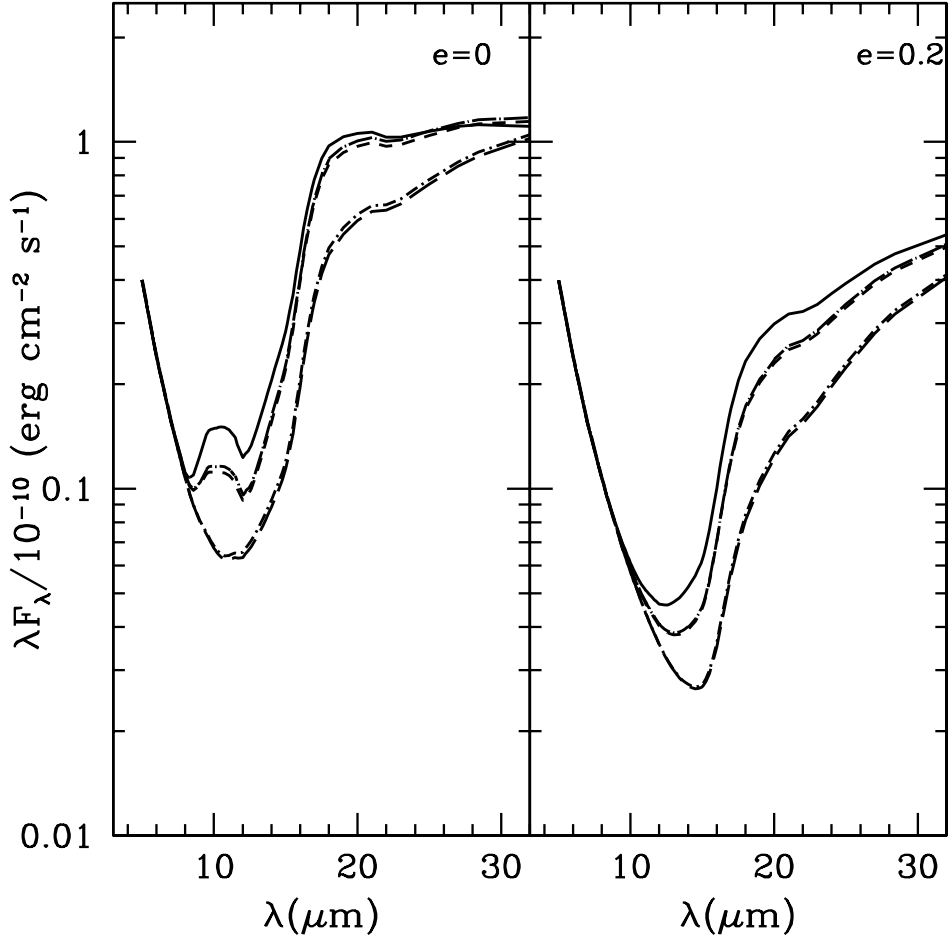}
\caption{Variation of the synthetic SED with $\alpha$ for $e=0$ (left plot) and
$e=0.2$ (right plot), with $\phi=0.5$,$\cos i=0.5$, and 
$h=0.28a$.  For the $e=0$ case, the models are: E1, $\alpha=0.0$ 
           (solid),AE1, $\alpha=0.25$ (small-dashed), AE2, 
	    $\alpha=0.5$ (long-dashed), AE3, 
	    $\alpha=1.5$ (dot-dashed) and AE4, 
	    $\alpha=1.75$ (dot-long dashed). 
            For the $e=0.2$ case, the models are: AE5, $\alpha=0.0$ 
           (solid),AE6, $\alpha=0.25$ (small-dashed), AE7, 
	    $\alpha=0.5$ (long-dashed), AE8, 
	    $\alpha=1.5$ (dot-dashed) and AE9, 
	    $\alpha=1.75$ (dot-long dashed). See table~\ref{table-models} 
for more details about the models.}
\label{fig-Draine-4Myr-alfa}
\end{figure}

Figure~\ref{fig-Draine-4Myr-alfa} shows the effects of changing the angle 
$\alpha$, for $\phi=0.5$,$\cos i=0.5$,$R_{cb}=1.7a$ and $h=0.28a$. 
Results are shown for $\alpha={0,0.25,0.5}$, ${1.5,1.75}$. 
In order to see the differences in terms of $e$, we show results for $e=0$ and 
$e=0.2$.
For symmetry arguments (with respect to the $\phi=0$ axis), the configurations for 
$\alpha=0.75$ and $1.25$ correspond to the same flux. The same can be said, for
the $\alpha=0.25$ and $1.75$ cases.
Moreover, the flux for all these cases is similar because the stars' 
masses are not so different and the binary orbit is circular. 
The configurations for $\alpha=0$ and $1$ correspond 
to the same flux, 
thus, the latter is not shown. The minimum flux is given in the $\alpha=0.5$
model, where the disk is inclined around a line perpendicular to the line that 
connects both stars, for $\alpha=0.5$ and $\alpha=1.5$ the $10\mu m$ band 
disappears. The maximum flux occurs in the $\alpha=0$ 
model (or $\alpha=1$). Changing $\alpha$ does not modify the area of the 
emitting region, since it only rotates this region in the plane of the sky.
However, since the stars are not at the center of the wall, 
when $\alpha$ changes the distances 
of both stars to the visible area of the wall also change, and thus the 
resulting SED changes. We can conclude
from Figure~\ref{fig-Draine-4Myr-alfa} that the cases are divided in two 
classes. The first one does not have a $10\mu m$ band and the second one has 
this feature. These models, where $\alpha$ changes, for cases $e=0$ and $e=2$
are shown in Table~\ref{table-models} as Series $\alpha,ecc$.  

\begin{figure}
\epsscale{.80}
\plotone{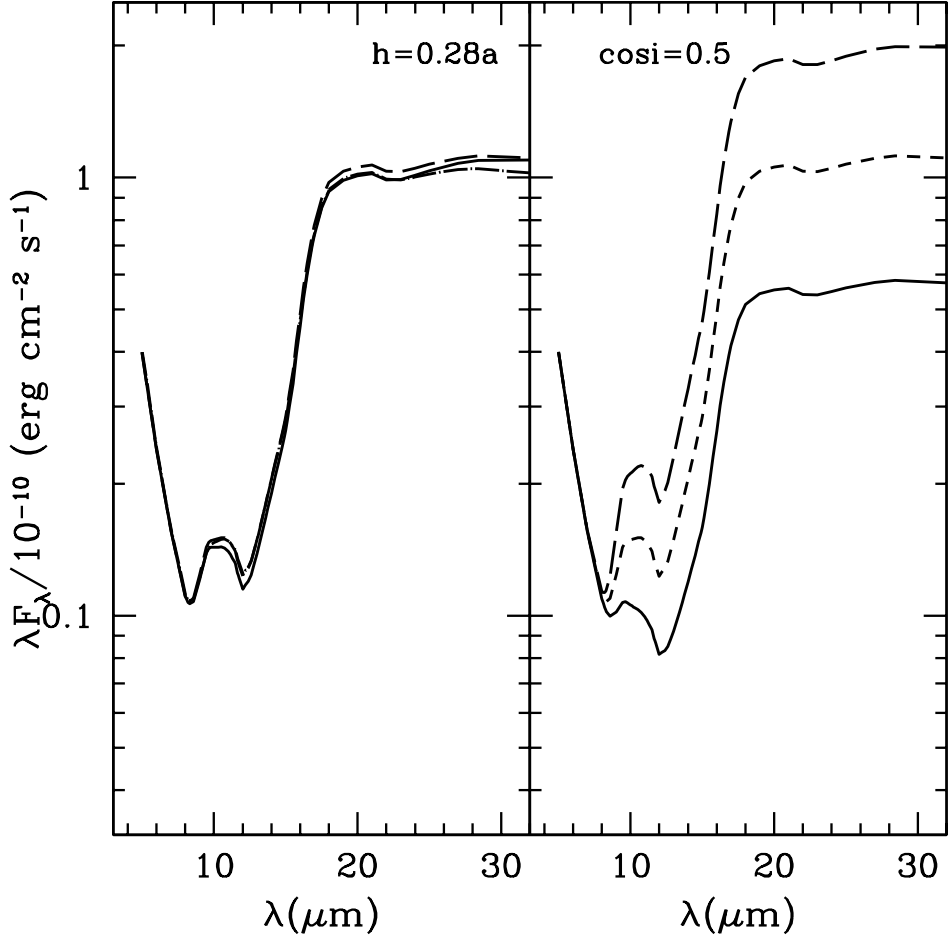}
\caption{The left plot illustrates the dependence of the synthetic SED with 
         inclination angle i, with $e=0$,$\phi=0.5$,$\alpha=0$,$R_{cb}=1.7a$. 
The plotted models are: IH1, $\cos i=0.3$
(solid), E1, $\cos i=0.5$ (long-dashed), and IH2, $\cos i=0.7$ 
	    (dot-long-dashed). The right plot show models for various $h$ with
the same values for $e$,$\phi$,$\alpha$ and $R_{cb}$.
            The models are: IH3, $h=0.14$ (solid), E1, $h=0.28$, (dashed)
            and IH4, $h=0.56$ (long-dashed). }
\label{fig-Draine-4Myr-i-h}
\end{figure}

Figure~\ref{fig-Draine-4Myr-i-h} shows the SED when $i$ and $h$ varies, for 
$e=0$,$\phi=0.5$,$\alpha=0$, and $R_{cb}=1.7a$. The 
left plot corresponds to $h=0.28a$ and $\cos{i}=0.3,0.5,0.7$, and the right one
is for $\cos{i}=0.5$ and $h=0.14,0.28,0.56$. The SED depends weakly on $i$, 
but, it has a strong dependence on $h$, namely, the flux increases for larger $h$. 
The reason for the former, is that for a perfectly vertical wall, the flux 
does not change monotonically with $i$. Of course, this effect will change if
a realistic shape of the wall is taken. However, at this moment there are no
studies (as far as we know) which define this shape (see 
Section~\ref{sec-conclusions}), thus, we assume a vertical wall.  Also, we are 
ignoring possible oscillation modes in the
wall \citep{Hayasaki} due to the tidal potential of the binary. If the
disk is viewed pole-on there is no visible area. In the opposite case, the 
emission from the wall of a disk seen edge-on is completely extinguished by
the disk itself.
Thus, the range in $\cos i$ relevant for modeling is small and 
the variation can be neglected.
In the present model, the height of the wall is taken as 
an arbitrary free parameter that can be adjusted to compensate for
any variation of $i$. Models showing changes in $i$ and $h$ are presented in
Table~\ref{table-models} as Series $i,h$. 


A complete analysis of all the possibilities that describe the dust composition
of the wall is computationally demanding, so we decided to explore the 
changes on the SED for typical cases. All these models can be used as a 
starting point to tackle future mid-IR spectra modeling with observations of 
new binary targets.
 
\begin{figure}
\epsscale{.80}
\plotone{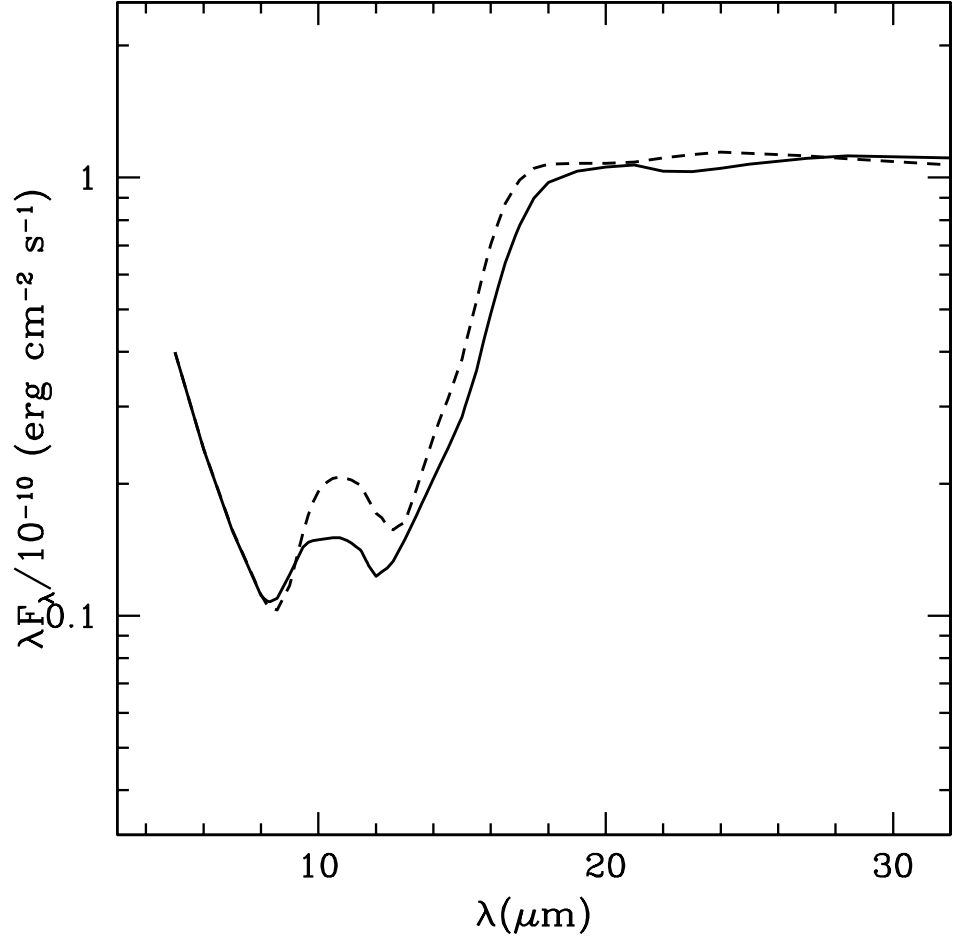}
\caption{Change of the SED for the fiducial model (E1) for two silicates: 
  pyroxene and olivines, with $e=0$,$\phi=0.5$,$\alpha=0$,$\cos i=0.5$,
$R_{cb}=1.7a$ and $h=0.28a$. The solid line corresponds to pyroxenes 
  (model E1), and the dashed line corresponds to olivines (model PO1).
   }
\label{fig-4Myr-pyr-oliv}
\end{figure}

Figure~\ref{fig-4Myr-pyr-oliv} shows the effect on the SED of changing the type
of silicates, for $e=0$,$\phi=0.5$,$\alpha=0$,$\cos i=0.5$,$R_{cb}=1.7a$ and 
$h=0.28a$. We show results for the fiducial model, and for the case where
the silicates component are changed from pyroxenes 
($Mg_{0.8}\,Fe_{0.2}\,SiO_{3}$) to olivines ($Mg\,Fe\,SiO_{4}$), model PO1 in
Table ~\ref{table-models} and the only member of Series pyr,oliv. 
The flux from the wall is larger when the dust is made by olivines (model P01)
than when it is made by pyroxenes (fiducial model). This can be explained,
because the former model has a higher temperature at the same distance from 
the central star. Another difference is that the central wavelength of the 
$10\mu$m feature due to olivines is slightly shifted to longer wavelengths
than the feature for the pyroxenes.

\begin{figure}
\epsscale{.80}
\plotone{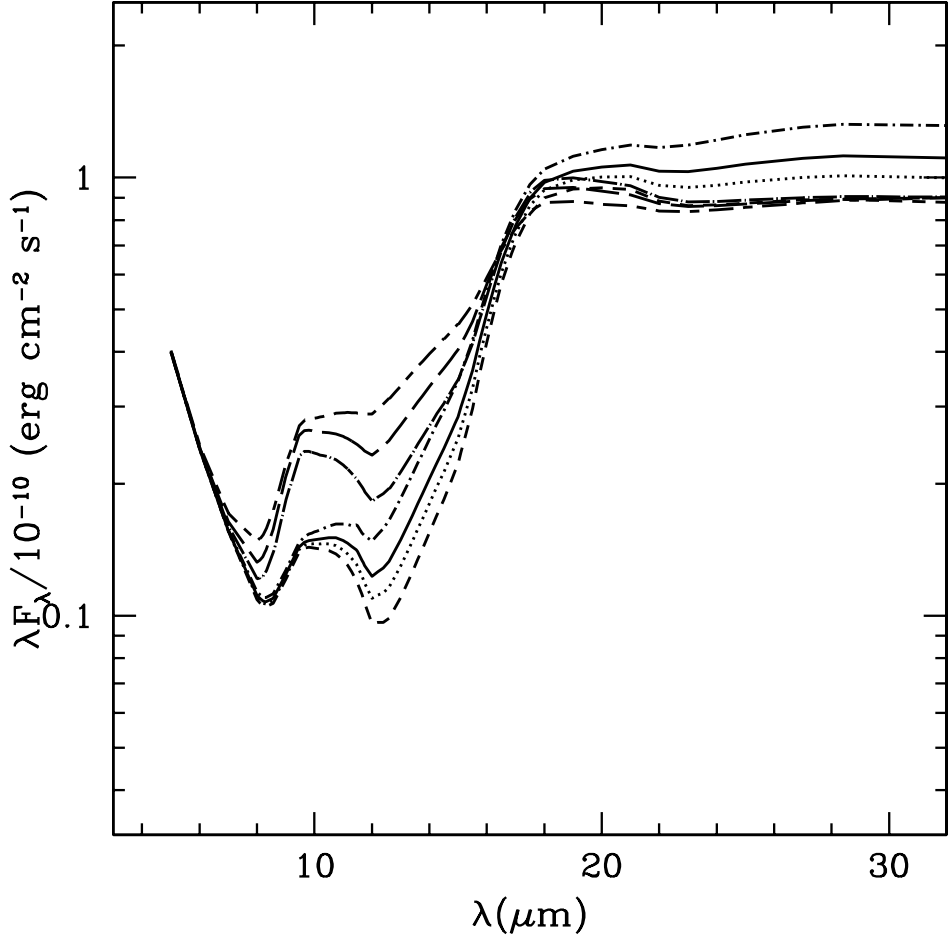}
\caption{Change of the SED for the fiducial model (E1) for various carbon
  components, with $e=0$,$\phi=0.5$,$\alpha=0$,$\cos i=0.5$,$R_{cb}=1.7a$ and 
  $h=0.28a$. The solid line corresponds to the fiducial model (E1,
  $\zeta_{grap}=0.0025$), the dashed
  line corresponds to the model with zero graphite abundance (PA1), and the
  long-dashed curve shows the model with amorphous carbon instead of graphite 
  with the same abundance (PA2). Also, we show cases with zero amorphous 
  carbon abundance for $\zeta_{grap}=0.00125$ (PA3,dotted), and for 
  $\zeta_{grap}=0.0050$ (PA4,dot-dashed). Other cases with zero graphite
  abundance include $\zeta_{am}=0.00125$ (PA5,dot-long-dashed) and
  $\zeta_{am}=0.0050$ (PA6,small-long-dashed).}
\label{fig-4Myr-pyr-graf-carb}
\end{figure}

Figure~\ref{fig-4Myr-pyr-graf-carb} shows the effect on the SED of changing the
type of carbon, for $e=0$,$\phi=0.5$,$\alpha=0$,$\cos i=0.5$,$R_{cb}=1.7a$ and 
$h=0.28a$. We show results for the case where we take amorphous carbon
\citep{Mathis3} and graphite \citep{Weingartner} grains. 
First, we compare two cases: a model with no graphite nor amorphous carbon
(i.e., only pyroxene grains in the wall, model PA1) and a model like the 
fiducial one, but where graphite has been replaced by amorphous carbon (model
PA2). We can see that model PA1 has
a lower flux than model PA2,
due to a lower 
$T_{wall}$, as noted in Table ~\ref{table-models}. The presence of amorphous 
carbon in model PA2 results in a larger temperature. 
For the same abundances, the absorption coefficients for graphite and 
amorphous carbon are similar between 0.3 and 3 microns, with the 
graphite opacity slightly larger. Outside this interval, the amorphous 
carbon opacity is higher.
Thus, changes in the SED (see Figure~\ref{fig-4Myr-pyr-graf-carb})
are expected due to the differences in the opacity for both components. 
 
Also in Figure~\ref{fig-4Myr-pyr-graf-carb}, models with variations on the 
abundance of graphite and amorphous 
carbon are also analyzed, for $e=0$,$\phi=0.5$,$\alpha=0$,$\cos i=0.5$,
$R_{cb}=1.7a$ and $h=0.28a$. Results are shown for the cases where there is no 
amorphous
carbon and the graphite abundance is halved and doubled (models PA3 and PA4 in
table ~\ref{table-models}). In another case, the graphite 
is replaced by amorphous carbon; model PA5 corresponds to half the adopted
standard value for amorphous carbon abundance and PA6 to the doubled value. 
For graphite and amorphous carbon, the flux increases as $\zeta$ 
increases. Model PA1 has the lowest 
flux. Physically this implies that silicates alone are heated at a much lower 
temperatures than the fiducial model, for the adopted value of $R_{cb}$.
All the PA models are assembled in Table~\ref{table-models} as Series
pyr,amorph.
  
\begin{figure}
\epsscale{.80}
\plotone{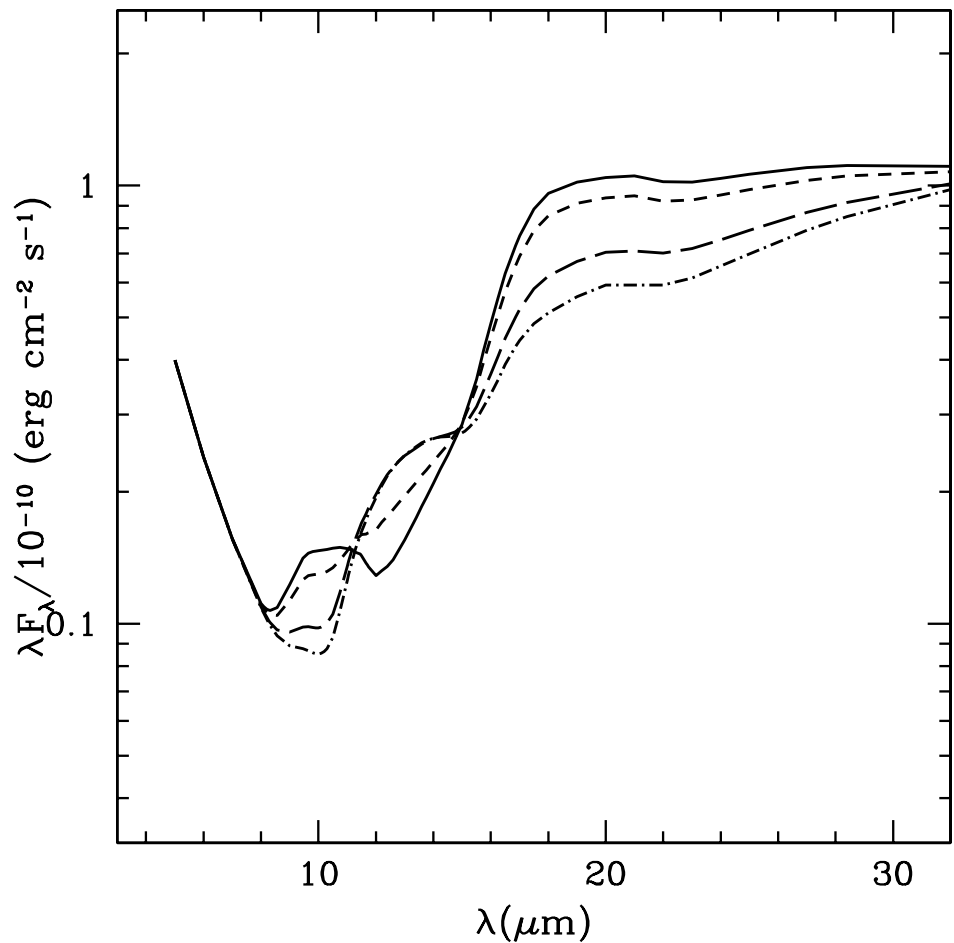}
\caption{Change of the SED for the fiducial model (E1), including water ice, 
 with $e=0$,$\phi=0.5$,$\alpha=0$,$\cos i=0.5$,$R_{cb}=1.7a$ and $h=0.28a$. 
  The plotted models are $\zeta_{ice}=0.000056$ (PI1,solid),
  $\zeta_{ice}=0.00056$ (PI2,dashed),$\zeta_{ice}=0.0028$ (PI3,long-dashed),
  and $\zeta_{ice}=0.0056$ (PI4,dot-dashed).  
  }
\label{fig-4Myr-pyr-graf-hielo}
\end{figure}

Figure~\ref{fig-4Myr-pyr-graf-hielo} shows the effect on the SED of changing 
the water ice abundance, for $e=0$,$\phi=0.5$,$\alpha=0$,$\cos i=0.5$,
$R_{cb}=1.7a$ and $h=0.28a$. The results are for four models, the water ice to
gas mass ratios are $\zeta_{ice}=0.0056$ (model PI4, abundance proposed by 
\citet{Pollack} for accretion disks), $50\%$ (model PI3), $10\%$ (model PI2) 
and $1\%$
(model PI1) of the Pollack et al. value. These models are labeled in 
Table~\ref{table-models} as Series pyr,ice.  We also use optical properties for
crystalline water ice given by \citet{Warren}. 
We can see that for the two higher
values of ice abundances, the $10\mu$m feature disappears. The $10\%$ case 
distorts this band significantly. Thus, for 
only $1\%$ of this abundance, the SED is similar to our fiducial model.
As we see in \S~\ref{sec-results} and \S~\ref{sec-conclusions}, this 
constrains the maximum allowed abundance of ice in the case of CoKu Tau/4.
The wall temperature decreases as the ice abundance increases, thus, the 
presence of the $10\mu$m band critically depends on $T_{wall}$. 



\begin{figure}
\epsscale{.80}
\plotone{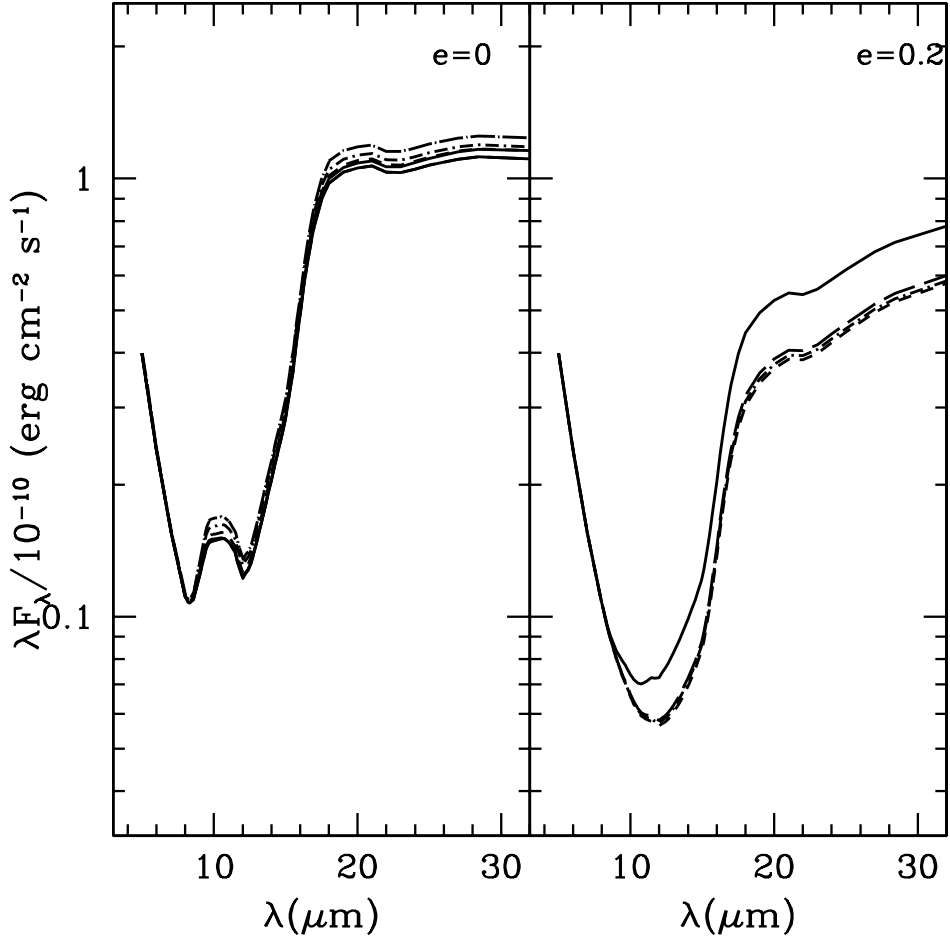}
\caption{Time evolution of the synthetic SED of an $e=0$ (left plot) and an 
         $e=0.2$ (right plot) system. For the $e=0$ case, the plotted models 
         are: PE1, $t=0.0T/2$ (solid), PE2, $t=0.25T/2$ (small-dashed), 
         E1, $t=0.5T/2$ (long-dashed), PE3, $t=0.75T/2$ 
	    (dot-dashed), and PE4, $t=1.0T/2$ 
	    (dot-long dashed). For the $e=0.2$ case, the plotted models are:
	 PE4, $t=0.0T/2$ (solid), PE5, $t=0.2T/2$ (small-dashed), 
         PE6, $t=0.45T/2$ (long-dashed), PE7, $t=0.7T/2$ 
	    (dot-dashed), and PE8, $t=0.9T/2$ (dot-long dashed). T is the 
         orbital period. For both plots, $\alpha=0$,$\cos i=0.5$, and $h=0.28a$}
\label{fig-Draine-4Myr-t}
\end{figure}

Finally, Figure~\ref{fig-Draine-4Myr-t} shows the effect on the SED at 
different times, for $\alpha=0$,$\cos i=0.5$, and 
$h=0.28a$. The results are shown for the evolution of the fiducial 
model in half an orbit, for times $t=$
$\phi\times T/2={(0,0.25,0.5,0.75,1.0)}\times T/2$, where $T$ is the orbital 
period.
The maximum variation in the spectra occurs between $t=0$ and $t=T/2$. In
order to complete the orbit, the sequence of plots must be followed in inverse 
order. As expected, a circular binary system inside a circular CB disk 
shows small changes, within 10 \%. Also in Figure~\ref{fig-Draine-4Myr-t}, the 
time variation for an $e=0.2$ system is shown. The results presented are for 
$t={(0,0.2,0.45,0.7,0.9)}\times T/2$. When the stars are along a line 
perpendicular to the major axis, the wall emits its largest observable
flux. The spread in fluxes is great given that in an eccentric system, the 
distances between the stars and the wall differ widely along the orbit. 
All these cases are 
placed under the name Series t,ecc in Table~\ref{table-models}.


\subsection{Effects of dust in the hole}
\label{sec-gap-models}



\begin{figure}
\epsscale{.80}
\plotone{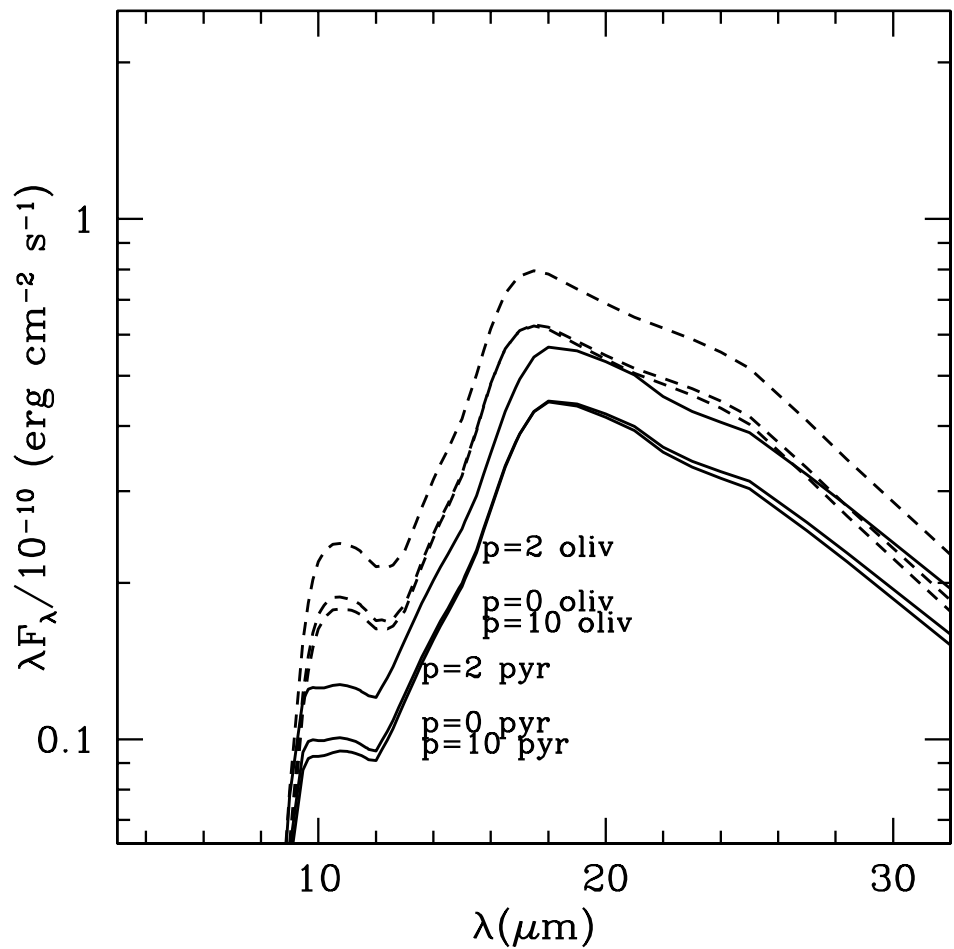}
\caption{Emission of optically thin material in the disk hole of
a $4 \ Myrs$ old binary system.  Pyroxenes: $p=0$ (middle),$p=2$ (upper),
   $p=10$ (lower), all solid lines.
   Olivines: $p=0$ (middle),$p=2$ (upper),$p=10$ (lower), all dashed lines. 
   For each case, the mass of dust in the hole is $1\times 10^{-9}M_{\odot}$
and it is distributed between R=1.4a and R=1.7a.}
\label{fig-4Myr-thin-p}
\end{figure}

Figure~\ref{fig-4Myr-thin-p} shows the 
differences in the shape of the SED produced by the material within the hole.
We present different values for the exponent of the mass surface density, $p=0,2,10$, in
the case of $e=0$. In this example, the 
optically thin material is assumed to be between 
 $R=R_{min}=1.4a$ and $R_{cb}=1.7a$ and is composed of $1\times 10^{-9}M_{\odot}$ of pyroxene or olivine.

The emission
of material highly concentrated in the wall ($p=10$) is smaller than  
 for the flatter density distributions ($p=0$ and $p=2$) for each type 
of silicates. 
Also, there
is considerably more emission for the hole filled with olivines than 
for the hole filled by pyroxenes.
The other noteworthy difference is that in the case of olivines, 
a $10\mu m$ band is more conspicuous in the SED than for pyroxenes.
The reason for all this is that, for the same system configuration,
the olivines temperature is larger than the pyroxenes' temperature. In addition, the 
band for olivine peaks at a longer wavelength ($~0.5\mu$m greater) than the pyroxene band, 
making it look more like a peak since this wavelength is in the Wien's
law side of the planck function.

\begin{figure}
\epsscale{.80}
\plotone{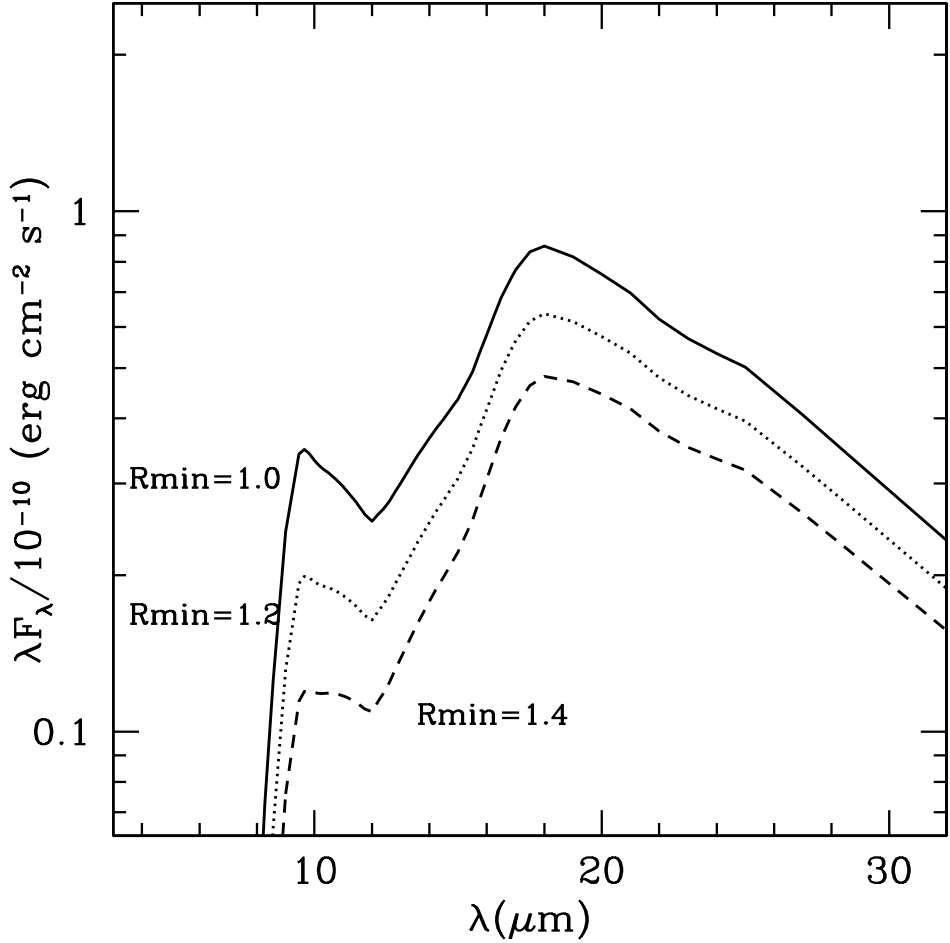}
\caption{Emission of optically thin material in the disk hole of
a $4 \ Myrs$ old binary system for various $R_{min}$. $R_{min}=1.0a$ (solid),
   $R_{min}=1.2a$ (dotted) and $R_{min}=1.4a$ (dashed). For each case, the
   mass of dust in the hole is $8\times 10^{-9}M_{\odot}$,
   $6.2\times 10^{-9}M_{\odot}$ and $4.8\times 10^{-9}M_{\odot}$ respectively,
   and it is distributed between $R=R_{min}$ and $R=1.7a$.}
\label{fig-4Myr-thin-Rmin}
\end{figure}

Figure~\ref{fig-4Myr-thin-Rmin} shows the effect on the SED of changing 
$R_{min}$, the inner 
boundary of material in the hole. We show results
for 3 cases: $R_{min}=1.0a$,$R_{min}=1.2a$ and $R_{min}=1.4a$. As expected, the 
flux is larger for the smaller value for $R_{min}$, for which the material lies 
in a larger region of the hole. 

\begin{figure}
\epsscale{.80}
\plotone{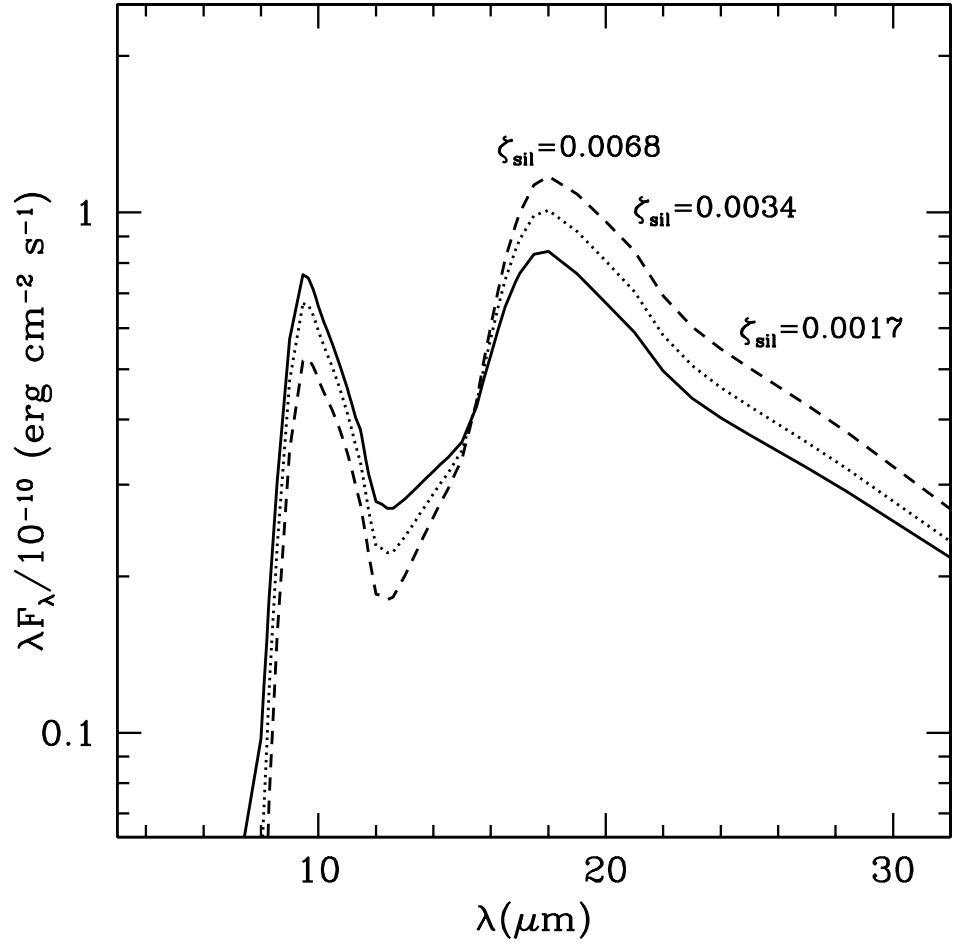}
\caption{Emission of optically thin material in the disk hole of
a $4 \ Myrs$ old binary system for various silicates abundances. 
$\zeta_{sil}=0.0017$ (solid), 
$\zeta_{sil}=0.0034$ (dotted) and $\zeta_{sil}=0.0068$ (dashed). For each case, 
the mass of dust in the hole is distributed between R=1.4a and R=1.7a.}
\label{fig-4Myr-thin-abund}
\end{figure}

Figure~\ref{fig-4Myr-thin-abund} shows the effect on the SED for different
abundances of silicates.
The results are for the case where graphite is included with the fiducial 
model abundance 
($\zeta_{grap}=0.0025$, \citet{Draine2}). The three studied cases are the 
standard value taken from \citet{Draine2} ($\zeta_{sil}=0.0034$) as well as 
half and twice this value. Note that diminishing $\zeta_{sil}$ does not
imply that the flux is smaller at all wavelengths because the opacity decreases,
but the dust temperature increases. 
Changing $\zeta_{sil}$ and keeping $\zeta_{grap}$ constant means that we are
changing the total dust to gas mass ratio and also the relative abundances of
silicate and graphite. In this case both the fluxes and the shape of the SED 
change. On the other hand, since the emission emerges from an optically thin 
region, changing the dust to gas mass ratio while keeping the relative
abundances of each ingredient constant, will make the shape of the SED remain 
the same. In this case, the fluxes will scale proportionally with the dust mass
inside the hole. Thus, the inferred dust mass is sensitive to the dust 
properties and composition.

\section{Application to the SED of CoKu Tau/4}
\label{sec-results}

\begin{figure}
\epsscale{.80}
\plotone{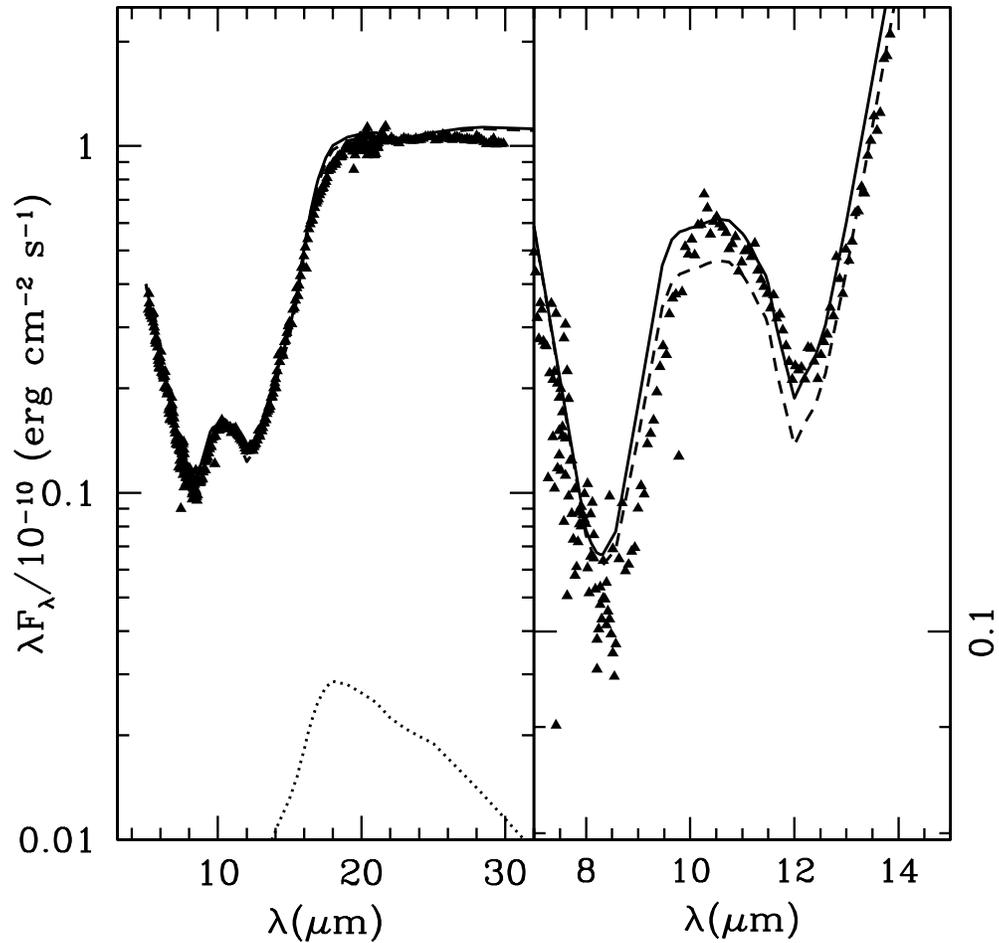}
\caption{SED of model M1, with $e=0$, age$= 4 \ Myrs$ (see table
~\ref{table-models}) compared to the
Spitzer IRS SED of CoKu Tau/4 (points), corrected with the reddening
	    law of \protect\citet{Draine1}. We show the total SED (solid line), the
            contribution from the wall (dashed line) and from
            optically thin material in the hole (dotted line). The right plot
            shows the details of the $10\mu$m band.}  
\label{fig-Draine-4Myr}
\end{figure}

The fiducial model was made using the known properties of CoKu Tau/4, and the 
geometrical restrictions given by the dynamical interaction between the binary
system and the disk \citep{Pichardo1,Pichardo2,Artymowicz1}.
For the latter issue, note that the D05 models required a wall radius between 
$9$ and $12AU$. For instance, $e=0.8$, $R_{cb}\approx 3.5 a$
\citep{Pichardo1}, and $a\sim 8$ AU \citep{Ireland} 
implies $R_{cb} \sim 28$ AU, i.e., three times the required value in D05.
Even for $e=0$, \citet{Pichardo1} find $R_{cb}\approx 1.9a$, giving 
$R_{cb} \sim 15$ AU, still too big for the implicitly required 
temperatures to explain the observed SED. 
However, including the disk viscosity,  \citet{Artymowicz1}  
find that for $e=0$,  $R_{cb}\approx 1.7a$, which corresponds 
to $\sim 13.5$ AU, still larger, but much closer to the wall radius estimated 
in D05. In this section, we show that the value for $R_{cb}$ given in 
\citet{Artymowicz1} results in a model of CoKu Tau/4, that fits the 
observed SED, as shown in Figure~\ref{fig-Draine-4Myr}.

In Figure~\ref{fig-Draine-4Myr-exc} one can see that
for $e>0$ the wall is too cold, because the $10\mu$m band does not match the
observations. 
 Even in some cases there is no band at $10\mu m$.    
Thus, a nearly circular binary orbit seems to be more consistent with the 
observed SED.
We present a model, for a  $4 \ Myrs$ system (model M1 in 
table~\ref{table-models}) in Figure~\ref{fig-Draine-4Myr} for CoKu Tau/4
(labeled Series CoKu Tau/4), where only the parameters of the wall are presented.
Indeed, this corresponds to the fiducial model (model E1) but here it is
repeated for clarity.
The spectral types and masses of the stars consistent with the observed
spectrum at high frequencies are $M1$ and $0.5M_\odot$ and
$M0$ and $0.6M_\odot$, respectively. The spectrum
corrected using the Draine's law is better fitted by 
pyroxene dust grains. The higher wall temperature when olivines are used
produces a flux larger than observed. Figure~\ref{fig-stars-spectrum} 
presents the observations dereddened with \citet{Draine1}, \citet{Mathis2},
\citet{McClure} and \citet{Moneti} laws. The flux for the observations 
dereddened with the \citet{Moneti} law is larger than that using the 
\citet{Draine1} law, the latter being taken here. Thus, olivine is a good 
option 
for modeling CoKu Tau/4, when using the Moneti law. A change in $h$ can also 
increase the flux, but the depth of the $10\mu$m feature changes 
(see Figure~\ref{fig-Draine-4Myr-i-h}). The spectrum dereddened with the 
\citet{McClure} law using $A_{V}=2.5$ is indistinguishable from the data 
dereddened with 
\citeauthor{Draine1}'s law and $A_{V}=3$. Thus, the model presented here can be 
used for both cases.
We use $h=0.28a$ ($h=0.15R_{cb}$) as a free parameter adjusted to fit the 
observed SEDs, but in 
the future it would be good to have an estimated range of $h$ consistent with
hydrodynamical arguments.  
Also, the wall cannot have a larger abundance of water ice 
($\zeta_{ice}>5.6\times 10^{5}$); otherwise, it would change the shape of the 
10 $\mu$m observed band, as described in \S~\ref{sec-var-param-wall} and
shown in Figure~\ref{fig-4Myr-pyr-graf-hielo}. From 
Figure~\ref{fig-4Myr-pyr-graf-carb} one can conclude that amorphous carbon
instead of graphite does not help to improve the fit,
because the resulting $10\mu$m band does not match the  observed SED.


In the case of CoKu Tau/4, there is no
evidence for circumstellar disks; thus, only a very small amount of mass 
should be able to cross the
inner boundary of the CB disk. But even so, it could be enough material 
to produce a contribution to the mid-IR SED. 
Our wall emission model for CoKu Tau/4 shown in Figure~\ref{fig-Draine-4Myr} 
seems to produce a reasonable fit to the observed SED. This fit can be improved
if optically thin material 
in the hole should be included. The fiducial model requires more flux at the
right side of the $10\mu$m band. Thus, looking at the inner hole emission models
in \S~\ref{sec-gap-models}, we choose the one with $p=1$, $R_{min}=1.4a$ 
and pyroxenes dust grains (see Figure~\ref{fig-4Myr-thin-p}). A model with a 
lower 
$R_{min}$ or with graphite has a prominent band at $10\mu$m, which implies a 
contribution at both sides of the fiducial model band, which is not what is 
needed in this case. The surface density profile is characterized by an 
exponent $p=1$ (see~\S~\ref{sec-gap-emission}). The total mass in dust 
in this region is $0.006$ lunar masses. 
The importance of this final fit is that the 
optically thin material lies outside the Lagrangian point radius ($R_{L_2}$),
i.e. the particles are located in a dynamically allowed region.
Some of these particles will be able to accrete onto 
the stars due to viscosity effects \citep{Artymowicz1,Artymowicz2} or
due to a not well-determined interior
boundary ($R_{min}$). However, even if this mass is accreted, and a
standard dust-to-gas mass ratio is assumed, the mass accretion rate onto the 
stars is below the observational threshold, $\dot{M} < 10^{-12}M_{\odot}/year$ 
\citep{Muzerolle}.  

Figure~\ref{fig-Draine-4Myr} shows the 
model with and without material in the hole and the contribution from the 
optically thin material.
Thus, the best-fit model to the emission of CoKu Tau/4 is the fiducial model,
plus optically thin material distributed inside a geometrically thin ring close
to the inner circumbinary wall.

\section{Conclusions}
\label{sec-conclusions}

We found a model that explains the observed mid-IR SED of CoKu Tau/4.
The model has a binary system in a nearly circular orbit with $4 \ Myrs$ stars 
old (using the evolutionary tracks from \citet{Siess}) and 
a circumbinary disk truncated at $R_{cb}=1.7 a \sim 12.6$ AU, with $a=7.43$ AU.
The height of the inner wall of the CB disk is around $h/a \sim 0.3$, for 
$i=60 ^\circ$. 
Also, we find that the inner hole might have a very small amount of dust 
($\lesssim 0.006$ lunar masses) with 
a spatial distribution in agreement with dynamical arguments of the binary-disk
interaction (see \citet{Artymowicz1}, \citet{Pichardo1} and \citet{Pichardo2}).
A detailed study of the amount and distribution of mass in the hole, can give us
information about the viscously driven process that allows for some particles
to fall into the hole when loosing angular momentum. This phenomenon strongly
depends on $h$ and $\alpha$ (viscosity coefficient from $\alpha$-disk models,
\citet{Shakura}), thus such a study can indirectly characterize these
parameters. The model for the optically thin material in the hole can be used
for arbitrary distributions of material in order to study the streams. The
structure of the streams can be taken from the hydrodynamical simulations in 
\citet{Gunther1} and \citet{Gunther2}. However, in this preliminary study we 
adopt an axisymmetric dust 
distribution which might represent some kind of average of these streams'models.
We leave a more detailed approach for future work.
 
Being a binary in a circular orbit,
we predict a small variability ($\sim$ 10 \%) of the SED of CoKu Tau/4 
in half its orbital timescale ($\sim$ 10 years),   
related to the changing distances between the stars and the 
CB disk. Other sources of variability are not accounted for in the
present model. This prediction can be tested if observations at other times are
available for this or other binary systems. On the other hand, 
if the binary orbit is not circular we found that no model can explain  the
observed SED because the inner wall of the CB disk is too far away from the
stars to have the appropriate temperature.

The study of the inner wall and the constraints on the dust sizes and
abundances allows us to probe the dust even at the midplane, which is
where planets must form. Note that Spitzer is sensitive to the dust located in 
the atmosphere of typical disks, not for matter in the midplane.
The wall can be seen as a perpendicular cut through the disk. Thus, the details
of the vertical structure are clearly exposed. In particular, the midplane is 
the place where material accumulates by settling. Eventually, the mass in this
region is 
above the gravitational instability threshold \citep{Goldreich}, and
planetesimals are able to form \citep{Youdin}. Thus, the study of this 
region is of prime importance. The advent of ALMA will allow us to actually see
the wall. The current interferometers are able to get a resolution of $1''$,
and at distances of typical regions of star formation, that means resolutions
$\sim 15AU$. ALMA will have five times better resolution \citep{Guilloteau1}, 
thus the region of optically thin material inside the wall and 
the location of the wall can be
resolved. The images produced can give us hints of the planetesimal formation
process which will result in a complementary study to the disk structure 
analysis as the
one presented here. Thus, studies of circumbinary and transitional disks, in 
particular the analysis of grain growth and settling in the wall, can place these 
disks in an evolutionary scheme. The dust composition can be traced
since the formation process of the star+disk system, (due to the fact that much
of the crystallization,just to name one process, occurs in the first stages of
the formation, \citet{Dullemond2}), until the phase of very evolved
stars \citep{Gielen}. The connection between evolved stars and the 
current solar system is seen in binaries like AC Her and RU Cen. The infrared
emission from the CB post-AGB disk shows a strong resemblance to the spectrum 
of the 
solar system primitive comet Hale-Bopp \citep{Gielen}. Thus, studies of
dust emission of the inner part of the CB disk are important to improve our
understanding of the bridge between young and old disks, from the formation of 
the disk to the formation of a planetary system.

In general, the model is not restricted to any configuration nor
any particular combination of stars, so it could be applied and tested with 
other observed binaries with circumbinary disks.  
It is clear that observations with high resolution, like the
ones described by \citet{Ireland}, in which both 
stars can be characterized, are very useful. In the near future, other binary 
systems
observed with Spitzer: Hen 3-600 \citep{Uchida}, St 34 \citep{Hartmann}, 
HD 98800 \citep{Furlan}, will be studied with this model. For these
cases, the assumption was made that the binary can be approximated as an 
equivalent star with the luminosity of both. However, the results that
come from this work, argue that this approximation should be taken with 
care.
Finally, a more detailed wall structure can be adopted,
as the inner curved rim described by \citet{Isella} or \citet{Tannirkulam}. In 
these works, the hole was formed by dust sublimation and the 
shape of the wall was given in the former by a 
density dependent sublimation of grains, and in the latter by the vertical 
differentiation 
of grain sizes. In the case of a circumbinary disk, it is not so clear 
what the shape of the disk inner wall would be, but one expects that dynamical effects
of the stars, hydrodynamical instabilities, pressure, viscosity or even
evaporation of the disk due to radiation from the stars, will have a major
role in the final shape of the wall.
The relevance of the unrealistic assumption of a vertical wall is not the 
aim of this work, however a full characterization  
requires a detailed analysis, which is beyond the scope of this paper, 
and will be addressed in a following paper.  

\acknowledgments

This work is based on observations made with the {\it Spitzer Space Telescope},
which is operated by the Jet Propulsion Laboratory, California Institute of
Technology, under contract with NASA. We are indebted with Adam Kraus for the 
opportunity to exchange ideas with him.
PD acknowledges a grant from CONACyT, M\'exico. EN appreciates postdoctoral 
fellowships from UNAM and CONACyT, M\'exico.  
CE was supported by the National Science Foundation under Award 
No. 0901947.





\appendix
\section{APPENDIX}






The cooling/heating time is:

\begin{equation}
  t_{c/h} = {K\Sigma_{R}\over 2\mu m_{H} \sigma T^3},
\end{equation}

where $\Sigma_{R}=\int^{R_{cb}+dR}_{R_{cb}}\rho dR$. Note that we are
interested in the area parallel to the wall surface.
The flux is equal to $\sigma T^4$, assuming blackbody emission at
temperature $T$.

Now, a typical surface density for disks around T Tauri stars located at
$10AU$ is $\Sigma_{d}\approx 10gcm^{-2}$ \citep{Dalessio1}, and the wall
height in the fiducial model is $H=0.28a=3.34\times 10^{13}cm$. Thus, an 
estimate for the volumetric density is

\begin{equation}
  \rho={\Sigma_{d}\over H}\approx 3\times 10^{-13}gcm^{-3}.
\end{equation}

The emission from the wall is optically thick, such that the emission comes
from a small radial region. In order to give an estimate for $t_{c/h}$, we 
take $\Delta R=H$, thus $\Sigma_{R}=\Sigma_{d}=10gcm^{-2}$. For the
temperature $T$, we take $T=<T_{wall}>=136.5K$, which corresponds to the
fiducial model. Thus, the cooling/heating time is:

\begin{equation}
  t_{c/h}\approx 2.86\times 10^{6}(\Sigma_{R}/10gcm^{-2})(T_{wall}/135.6)^{-3} 
         (\mu/1.)^{-1}s.
\end{equation}

The orbital time is simply

\begin{equation}
  t_{orb}={a^{3/2}\over (G(M_{\star 1}+M_{\star 2}))^{1/2}}=1.08\times 10^{8}s.
\end{equation}

We find that for typical parameters, $t_{c/h}<<t_{orb}$, justifying the 
assumption of an instantaneous adjustment of the disk inner edge to changing
stellar irradiation.

{}

\clearpage

\clearpage

\clearpage

\clearpage


\clearpage

\clearpage

\clearpage


\clearpage


\clearpage

\clearpage

\clearpage


\clearpage


\clearpage

\clearpage

\clearpage

\clearpage

\clearpage

\clearpage

\clearpage

\clearpage

\clearpage



\clearpage

\begin{deluxetable}{ccccc}
\rotate
\tablewidth{0pt}
\tablecaption{Stellar Parameters\label{table-stars}}
\tablehead{
\colhead{spectral type}  &
\colhead{$M_{\star}(M_{\odot})$}      &
\colhead{$T_{\star}(K)$}     & 
\colhead{$R_{\star}(R_{\odot})$}  &
\colhead{$L_{\star}(L_{\odot})$}}
\startdata
M1 & 0.5 & 3894 & 1.178 & 0.2886 \\
M0 & 0.6 & 4027 & 1.261 & 0.3782  
\enddata
\end{deluxetable}

\clearpage

\begin{deluxetable}{cccccccccccc}
\tabletypesize{\scriptsize}
\rotate
\tablecaption{Models discussed in this paper\tablenotemark{a}\label{table-models}}
\tablewidth{0pt}
\tablehead{
\colhead{Series}       & \colhead{Figure}      &
\colhead{Model}        & \colhead{$a(AU)$}    &
\colhead{$e$}          & \colhead{$\phi(\pi)$}  &
\colhead{$\alpha(\pi)$}     & \colhead{$\cos i$} &
\colhead{$R_{cb}(a)$}  & \colhead{$h(a)$}    &
\colhead{Dust\tablenotemark{b}}   &
\colhead{$T_{wall}$} 
}
\startdata
Ecc & 5 & E1 & 7.43 & 0 & 0.5 & 0 & 0.5 & 1.7 & 0.28 & 1
 & 136.5  \\
\nodata & 5 & E2 & 9.287 & 0.2 & 0.5 & 0 & 0.5 & 2.35 & 0.28 & 1
 & 110.7  \\
\nodata & 5 & E3 & 12.38 & 0.4 & 0.5 & 0 & 0.5 & 2.7 & 0.28 & 1
 & 95.2  \\
\nodata & 5 & E4 & 18.575 & 0.6 & 0.5 & 0 & 0.5 & 2.95 & 0.28 & 1
 & 80.7  \\
$\alpha,ecc$ & 6 & AE1 & 9.40 & 0 & 0.5 & 0.25 & 0.5 & 1.7 & 0.28 & 1
 & 125.0  \\
\nodata & 6 & AE2 & 14.87 & 0 & 0.5 & 0.5 & 0.5 & 1.7 & 0.28 & 1
 & 103.0  \\
\nodata & 6 & AE3 & 14.87 & 0 & 0.5 & 1.5 & 0.5 & 1.7 & 0.28 & 1
 & 103.8  \\
\nodata & 6 & AE4 & 9.40 & 0 & 0.5 & 1.75 & 0.5 & 1.7 & 0.28 & 1
 & 125.6  \\
\nodata & 6 & AE5 & 9.29 & 0.2 & 0.5 & 0.0 & 0.5 & 2.9 & 0.28 & 1
 & 99.60  \\
\nodata & 6 & AE6 & 11.75 & 0.2 & 0.5 & 0.25 & 0.5 & 2.9 & 0.28 & 1
 & 91.21  \\
\nodata & 6 & AE7 & 18.58 & 0.2 & 0.5 & 0.5 & 0.5 & 2.9 & 0.28 & 1
 & 77.12  \\
\nodata & 6 & AE8 & 18.58 & 0.2 & 0.5 & 1.5 & 0.5 & 2.9 & 0.28 & 1
 & 77.44  \\
\nodata & 6 & AE9 & 11.75 & 0.2 & 0.5 & 1.75 & 0.5 & 2.9 & 0.28 & 1
 & 91.47  \\
$i,h$ & 7 & IH1 & 7.43 & 0 & 0.5 & 0 & 0.3 & 1.7 & 0.28 & 1
 & 136.1  \\
\nodata & 7 & IH2 & 7.43 & 0 & 0.5 & 0 & 0.7 & 1.7 & 0.28 & 1
 & 136.7  \\
\nodata & 7 & IH3 & 7.43 & 0 & 0.5 & 0 & 0.5 & 1.7 & 0.14 & 1
 & 136.9  \\
\nodata & 7 & IH4 & 7.43 & 0 & 0.5 & 0 & 0.5 & 1.7 & 0.56 & 1
 & 135.2  \\
pyr,oliv & 8 & PO1 & 7.43 & 0 & 0.5 & 0 & 0.5 & 1.7 & 0.28 & 1a
 & 146.2  \\
pyr,amorph & 9 & PA1 & 7.43 & 0 & 0.5 & 0 & 0.5 & 1.7 & 0.28 & 2
 & 136.5  \\
\nodata & 9 & PA2 & 7.43 & 0 & 0.5 & 0 & 0.5 & 1.7 & 0.28 & 3
 & 166.8  \\
\nodata & 9 & PA3 & 7.43 & 0 & 0.5 & 0 & 0.5 & 1.7 & 0.28 & 4
 & 136.5  \\
\nodata & 9 & PA4 & 7.43 & 0 & 0.5 & 0 & 0.5 & 1.7 & 0.28 & 5
 & 136.5  \\
\nodata & 9 & PA5 & 7.43 & 0 & 0.5 & 0 & 0.5 & 1.7 & 0.28 & 6
 & 158.8  \\
\nodata & 9 & PA6 & 7.43 & 0 & 0.5 & 0 & 0.5 & 1.7 & 0.28 & 7
 & 173.8  \\
pyr,ice & 10 & PI1 & 7.43 & 0 & 0.5 & 0 & 0.5 & 1.7 & 0.28 & 8
 & 136.2  \\
\nodata & 10 & PI2 & 7.43 & 0 & 0.5 & 0 & 0.5 & 1.7 & 0.28 & 9
 & 134.0  \\
\nodata & 10 & PI3 & 7.43 & 0 & 0.5 & 0 & 0.5 & 1.7 & 0.28 & 10
 & 128.3  \\
\nodata & 10 & PI4 & 7.43 & 0 & 0.5 & 0 & 0.5 & 1.7 & 0.28 & 11
 & 125.1  \\
$t,ecc$ & 11 & PE1 & 7.43 & 0 & 0 & 0 & 0.5 & 1.7 & 0.28 & 1
 & 135.8  \\
\nodata & 11 & PE2 & 7.43 & 0 & 0.25 & 0 & 0.5 & 1.7 & 0.28 & 1
 & 136.7  \\
\nodata & 11 & PE3 & 7.43 & 0 & 0.75 & 0 & 0.5 & 1.7 & 0.28 & 1
 & 137.3  \\
\nodata & 11 & PE4 & 7.43 & 0 & 1 & 0 & 0.5 & 1.7 & 0.28 & 1 
 & 137.9  \\
\nodata & 11 & PE5 & 9.29 & 0.2 & 0 & 0 & 0.5 & 2.9 & 0.28 & 1
 & 112.7  \\
\nodata & 11 & PE6 & 9.29 & 0.2 & 0.5 & 0 & 0.5 & 2.9 & 0.28 & 1
 & 108.3  \\
\nodata & 11 & PE7 & 9.29 & 0.2 & 1 & 0 & 0.5 & 2.9 & 0.28 & 1
 & 108.9  \\
\nodata & 11 & PE8 & 9.29 & 0.2 & 1.5 & 0 & 0.5 & 2.9 & 0.28 & 1 
 & 108.7  \\
CoKu Tau/4 & 15 & M1 & 7.43 & 0 & 0.5 & 0 & 0.5 & 1.7 & 0.28 & 1
 & 136.5  \\

\enddata
\tablenotetext{a}{The column titles are given as in the text.}
\tablenotetext{b}{The numbers in the column ``Dust''
refer to different adopted abundances. See 
Table~\ref{table-dust} for the dust abundances.}
\end{deluxetable}

\begin{deluxetable}{ccccccc}
\rotate
\tablewidth{0pt}
\tablecaption{Dust abundances\label{table-dust}}
\tablehead{
\colhead{Dust model}  &
\colhead{$\zeta_{grap}$}      &
\colhead{$\zeta_{sil}$}     & 
\colhead{$\zeta_{troi}$}  &
\colhead{$\zeta_{ice}$} & \colhead{$\zeta_{amorp}$} &
\colhead{wall sil\tablenotemark{a}}
}
\startdata
1 & 0.0025 & 0.0034 & 7.68E-4 & 0.0 & 0.0 & pyr \\
1a & 0.0025 & 0.0034 & 7.68E-4 & 0.0 & 0.0 & oliv \\
2 & 0.0 & 0.0034 & 7.68E-4 & 0.0 & 0.0 & pyr \\
3 & 0.0 & 0.0034 & 7.68E-4 & 0.0 & 0.0025 & pyr \\
4 & 0.00125 & 0.0034 & 7.68E-4 & 0.0 & 0.0 & pyr \\
5 & 0.005 & 0.0034 & 7.68E-4 & 0.0 & 0.0 & pyr \\
6 & 0.0 & 0.0034 & 7.68E-4 & 0.0 & 0.00125 & pyr \\
7 & 0.0 & 0.0034 & 7.68E-4 & 0.0 & 0.005 & pyr \\
8 & 0.0025 & 0.0034 & 7.68E-4 & 5.6E-5 & 0.0 & pyr \\
9 & 0.0025 & 0.0034 & 7.68E-4 & 5.6E-4 & 0.0 & pyr \\
10 & 0.0025 & 0.0034 & 7.68E-4 & 2.8E-3 & 0.0 & pyr \\
11 & 0.0025 & 0.0034 & 7.68E-4 & 0.0056 & 0.0 & pyr \\
\enddata
\tablenotetext{a}{''Wall sil'' refers to the type of silicates used in the wall.}
\end{deluxetable}




\end{document}